\documentclass{aastex63}

\usepackage{CJKutf8}
\newcommand{\TLchinesename}{{\begin{CJK}{UTF8}{gbsn}(李坦达)\end{CJK}}}
\newcommand{\YLchinesename}{{\begin{CJK}{UTF8}{gbsn}(李亚光)\end{CJK}}}
\newcommand{\SLchinesename}{{\begin{CJK}{UTF8}{gbsn}(毕少兰)\end{CJK}}}
\newcommand{\DMchinesename}{{\begin{CJK}{UTF8}{gbsn}(杜明昊)\end{CJK}}}
\makeatletter
\newcommand*\fsize{\f@size pt\relax}
\makeatother
\usepackage{comment}

\newcommand{\kepler}[0]{\emph{Kepler}}
\newcommand{\corot}[0]{\emph{CoRoT}}
\newcommand{\tess}[0]{\emph{TESS}}
\newcommand{\ktwo}[0]{\emph{K2}}
\newcommand{\apogee}[0]{\emph{APOGEE}}
\newcommand{\lamost}[0]{\emph{LAMOST}}

\newcommand{\gaia}[0]{\emph{Gaia}}
\newcommand{\teff}[0]{$T_{\text{eff}}$}

\newcommand{\Dnu}[0]{$\Delta\nu$}
\newcommand{\dnu}[1]{$\delta\nu_{#1}$}
\newcommand{\numax}[0]{$\nu_{\rm max}$}

\submitjournal{APJS}

\shorttitle{Asteroseismology of $\kepler$ Red Giants}
\shortauthors{Li et al.}
\graphicspath{{./}{figures/}}
\begin{document}

\title{Asteroseismology of 3,642 $\kepler$ Red Giants: Correcting the Scaling Relations based on Detailed Modeling}

\correspondingauthor{Tanda Li}
\email{t.li.2@bham.ac.uk}

\author[0000-0001-6396-2563]{Tanda Li\TLchinesename}
\affiliation{School of Physics and Astronomy, University of Birmingham, Edgbaston, Birmingham B15 2TT, UK}
\affiliation{Stellar Astrophysics Centre (SAC), Department of Physics and Astronomy, Aarhus University, Ny Munkegade 120, 8000 Aarhus C, Denmark}

\author[0000-0003-3020-4437]{Yaguang Li\YLchinesename}
\affiliation{Sydney Institute for Astronomy (SIfA), School of Physics, University of Sydney, NSW 2006, Australia}
\affiliation{Stellar Astrophysics Centre (SAC), Department of Physics and Astronomy, Aarhus University, Ny Munkegade 120, 8000 Aarhus C, Denmark}

\author[0000-0002-7642-7583]{Shaolan Bi\SLchinesename}
\affiliation{Department of Astronomy, Beijing Normal University, Beijing, 100875, China}

\author[0000-0001-5222-4661]{Timothy R. Bedding}
\affiliation{Sydney Institute for Astronomy (SIfA), School of Physics, University of Sydney, NSW 2006, Australia}
\affiliation{Stellar Astrophysics Centre (SAC), Department of Physics and Astronomy, Aarhus University, Ny Munkegade 120, 8000 Aarhus C, Denmark}

\author[0000-0002-4290-7351]{Guy Davies}
\affiliation{School of Physics and Astronomy, University of Birmingham, Edgbaston, Birmingham B15 2TT, UK}
\affiliation{Stellar Astrophysics Centre (SAC), Department of Physics and Astronomy, Aarhus University, Ny Munkegade 120, 8000 Aarhus C, Denmark}

\author[0000-0002-7642-7583]{Minghao Du\DMchinesename}
\affiliation{Department of Astronomy, Beijing Normal University, Beijing, 100875, China}



\begin{abstract}
The paper presents a correction to the scaling relations for red-giant stars using model-based masses and radii.
% We present the radial mode identifications and theoretical modeling for $\sim3,000$ solar-like oscillators on the red-giant branch observed by the $\it Kepler$ mission for the purpose of correcting the asteroseismic scaling relations.
We measure radial-mode frequencies from \kepler\ observations for 3,642 solar-like oscillators on the red-giant branch 
%observed by the $\it Kepler$ mission 
and use them to characterise the stars with the grid-based modeling.
We determine fundamental stellar parameters with good precision: the typical uncertainty is 4.5\% for mass, 16\% for age, 0.006 dex for surface gravity, and 1.7\% for radius. We also achieve good accuracy for estimated masses and radii, based on comparing with those determined for eclipsing binaries.  
We find a systematic offset of $\sim15\%$ in mass and $\sim7\%$ in radius between the modeling solutions and the scaling relations. Further investigation indicates that these offsets are mainly caused by a systematic bias in the $\Delta \nu$ scaling relation: the original scaling relation underestimates $\Delta\nu$ value by $\sim4\%$, on average, and it is important to correct for the surface term in the calibration. We find no significant offset in the $\nu_{\rm max}$ scaling relation, although a clear metallicity dependence is seen and we suggest including a metallicity term in the formulae. Lastly, we calibrate new scaling relations for red-giant stars based on observed global seismic parameters, spectroscopic effective temperatures and metallicities, and modeling-inferred masses and radii.  
\end{abstract}


\section{Introduction} \label{sec:intro}

Asteroseismology of red giants has been undergoing extensive development in the last decade, thanks to the space missions like \corot{}\citep{2002ESASP.485...17B}, \kepler \citep{2009IAUS..253..289B}, \ktwo{} \citep{2014PASP..126..398H} and \tess{} \citep{2015JATIS...1a4003R}. High-quality photometry data for thousands of red giants are now available for asteroseismology. Automated data reduction tools for measuring global seismic parameters \citep[e.g. SYD Pipeline;][]{2009CoAst.160...74H}, acoustic modes \citep[e.g. Pbjam;][]{2021AJ....161...62N}, and mixed modes \citep[e.g. FAMED;][]{2020A&A...640A.130C} make it possible to study solar-like oscillations in large samples of stars. Compared with studying individual stars, a population study covering a wide parameter range has great advantages. For instance, the systematic characterisation of solar-like oscillations and granulation can constrain stellar masses and radii with good precision \citep[e.g.][]{2018ApJS..236...42Y}. In addition, measuring gravity-mode period spacings for red giants allows us to precisely monitor their stellar evolution from the subgiant stage to the end of helium burning phase \citep{2014A&A...572L...5M,2016A&A...588A..87V}. Finally, core rotation periods, measured from the splitting in oscillation modes \citep[e.g.][]{2012A&A...548A..10M}, show a significant spinning down during the red-giant phase and provide constraints on the angular momentum transport between the core and the envelope \citep{2019ARA&A..57...35A,2019ApJ...887..203T}.

Studying a large sample of stars on the red-giant branch (RGB) could advance the resolution of puzzles at this evolutionary stage. One of these is the remarkable accuracy of the scaling relations. Scaling relations were introduced to predict the oscillation frequencies of solar-type stars \citep{1986ApJ...306L..37U,1991ApJ...368..599B,1995A&A...293...87K}:
\begin{equation}\label{eq:sc-Dnu}
   \frac{\Delta\nu}{\Delta\nu_{\odot}} \simeq \left (\frac{\rho}{\rho_{\odot}} \right)^{0.5} 
\end{equation}
and
\begin{equation}\label{eq:sc-numax}
   \frac{\nu_{\rm max}}{\nu_{\rm max,\odot}} \simeq \left (\frac{g}{g_{\odot}} \right) \left ( \frac{T_{\rm eff}}{T_{\rm eff, \odot}} \right )^{-0.5}.
\end{equation}
Here, $\nu_{\rm max}$ is the frequency at the maximum amplitude of stellar oscillations, and $\Delta\nu$ is the mean frequency separation between consecutive modes of the same spherical degree. These relations can be inverted to estimate masses and radii \citep{2008ApJ...674L..53S,2010A&A...509A..77K}:
\begin{equation}\label{eq:sc1}
   \frac{M_{}}{M_{\odot}} \simeq \left ( \frac{\nu_{\rm max}}{\nu_{\rm max,\odot}} \right )^{3}\left ( \frac{\Delta\nu}{\Delta\nu_{\odot}} \right )^{-4}\left ( \frac{T_{\rm eff}}{T_{\rm eff, \odot}} \right )^{1.5}, 
\end{equation}
 and
\begin{equation}\label{eq:sc2}
   \frac{R_{}}{R_{\odot}} \simeq \left ( \frac{\nu_{\rm max}}{\nu_{\rm max,\odot}} \right )\left ( \frac{\Delta\nu}{\Delta\nu_{\odot}} \right )^{-2}\left ( \frac{T_{\rm eff}}{T_{\rm eff, \odot}} \right )^{0.5}.
\end{equation}
The solar reference values have been measured in previous studies with slight differences, depending on the data and the reduction method. In this work, we adopt $\nu_{\rm max,\odot}$ = 3090$\pm$30$\mu$Hz, $\Delta\nu_{\odot}$ = 135.1$\pm$0.1$\mu$Hz, and $T_{\rm eff, \odot}$ = 5777K \citep{2011ApJ...743..143H}. 

The accuracy of scaling relations has been widely discussed \citep{2009ApJ...700.1589S,2011ApJ...743..161W,2012ApJ...760...32H,2013A&A...550A.126M,2021MNRAS.501.3162L} and noticeable systematic offsets have been seen in RGB stars.
Stars in eclipsing binary (EB) systems can provide direct tests of the scaling relations. 
For instance, \citet{2016ApJ...832..121G} studied ten high-luminosity red-giant stars in eclipsing binaries by comparing masses and radii obtained from the scaling relations against spectroscopic orbital solutions using dynamical modeling. They found that the scaling relations overestimate the radii by about 5\% and the masses by about 15\%. With the similar approaches, \citet{2018MNRAS.476.3729B} and \citet{2018MNRAS.478.4669T} also found overestimated masses and radii derived by the scaling relations.
The radius measured based on interferometry provides another direct test. For instance, \citet{2012ApJ...760...32H} combined interferometric angular diameters, Hipparcos parallaxes, asteroseismic densities, bolometric fluxes and high-resolution spectroscopy to derive near model-independent fundamental properties for one subgiant and four red giant stars near the base of RGB. Unlike the EB results, their measurements showed good agreement with the scaling relations.  
% It worth to note that overestimated temperatures can lead to overestimated masses and radii. When \citet{2016ApJ...832..121G} decrease effective temperatures by 100 K, the offsets decrease by 3.1\% for masses and 1.0\% for radii. 
% To test the scaling relation in a wide range of stellar parameters and evolution stages, 
% \citet{2019MNRAS.486.4612B} used masses, radii previously determined from fits to theoretical stellar models \citep{2015MNRAS.452.2127S,2016MNRAS.456.2183D, 2017ApJ...835..172L,2017ApJ...835..173S} for {\it Kpler} main-sequence stars.
% The sample covers a reasonable large area at the main-sequence and results showed that the scaling relations averagely overestimate masses and radii by 6.9\% and 2.4\%.

Determining masses and radii from the detailed modeling is another way to verify the scaling relations. \citet{2016A&A...591A..99P} characterised 19 low-luminosity \kepler{} red giants based on a minimisation procedure combining the frequencies of the p-modes, the period spacing of gravity modes, and the spectroscopic data. They found no obvious differences between inferred masses and radii from the modeling and the scaling relations. 
It should be noted that the apparent conflict between these various studies may relate to the evolutionary stage rather than differences in the methods.
As presented by \citet{2016ApJ...822...15S}, the systematic offset of the \Dnu{} scaling relation varies significantly for different evolutionary stages. 
Global diagnosis of the scaling relations that covers a wide parameter range gives a big picture of the systematic uncertainty in the scaling relations. \citet{2017ApJ...844..102H}, \citet{2019ApJ...885..166Z}, and \citet{2019MNRAS.486.3569H} used \gaia{} parallaxes to empirically demonstrate that evolution-dependent corrections to the scaling relations are necessary to improve the accuracy of asteroseismic radii for thousands of \kepler{} RGB stars.  
%
% However, we still lack a global diagnosis of the scaling relations that covers a wide parameter range on RGB and this is the goal of this work.

Various methods have been implemented to correct the scaling relations. A simple approach is changing the solar reference values of $\Delta\nu_{\odot}$ and $\nu_{\rm max,\odot}$ \citep{2010A&A...522A...1K,2011Sci...332..213C,2016MNRAS.460.4277G,2018MNRAS.478.4669T}, however, the method does not consider the change for different evolutionary phases. \citet{2016ApJ...822...15S}, \citet{2017MNRAS.467.1433R}, and \citet{2013A&A...550A.126M} provided a correction factor to $\Delta\nu$ ($f_{\Delta\nu}$), which is calculated based on the theoretical $\Delta\nu$ from fitting radial mode frequencies. This corrects the scaling relations for wide mass and metallicity ranges and also for different evolutionary phases, from the ZAMS to the upper RGB. 
% calculated the correction factor to $\Delta\nu$ ($f_{\Delta\nu}$) based on theoretical $\Delta\nu$ from fitting radial mode frequencies. 
Including these correction factors, the masses and radii reached general agreement for stars in the EB systems \citep[e.g.][]{2018MNRAS.476.3729B}. 
%
%
%Moreover, \citet{2018A&A...616A.104K} derived non-linear scaling relations based on six EBs and obtained a new $\nu_{\rm max}$ relation to the surface gravity and effective temperature, in which, they suggested that a curvature and glitch correction should be preferred over a local or average value of $\Delta\nu$. 
%
% \red{using mean weights $\mu$ to involve chemical info in the formulae}
More details about the efforts on this topic can be seen in the review by  \citet{2020FrASS...7....3H}. 

In this paper, we use masses and radii determined with the detailed modeling to correct the scaling relations. Although theoretical models do introduce potential systematic uncertainties, the advantage of using modeling solutions is the coverage of a wide range of stellar parameters and evolution stages. For instance, \citet{2019MNRAS.486.4612B} used masses and radii previously determined from fits to theoretical stellar models \citep{2015MNRAS.452.2127S,2016MNRAS.456.2183D, 2017ApJ...835..172L,2017ApJ...835..173S} to calibrate the scaling relations for $\kepler$ dwarfs.
We aim to apply this method to \kepler{} red-giant oscillators. We calculate radial model frequencies to determine stellar masses and radii and then use inferred values to calibrate the scaling relations. 
The rest of this paper is organised as follow. Section \ref{sec:obs} summarises the target selection and the measurement procedure of oscillating radial modes; Section \ref{sec:model} discusses our theoretical modeling and the fitting method; Section \ref{sec:results} demonstrates results of estimated stellar parameters; we correct the scaling relations with model-determined masses and radii in Section \ref{sec:scaling}; and we close with a summary of outcomes in Section \ref{sec:conclusion}.

\section{Data Reduction} \label{sec:obs}

\subsection{Target Selection}
\begin{figure}
\centering
\includegraphics[scale=0.32]{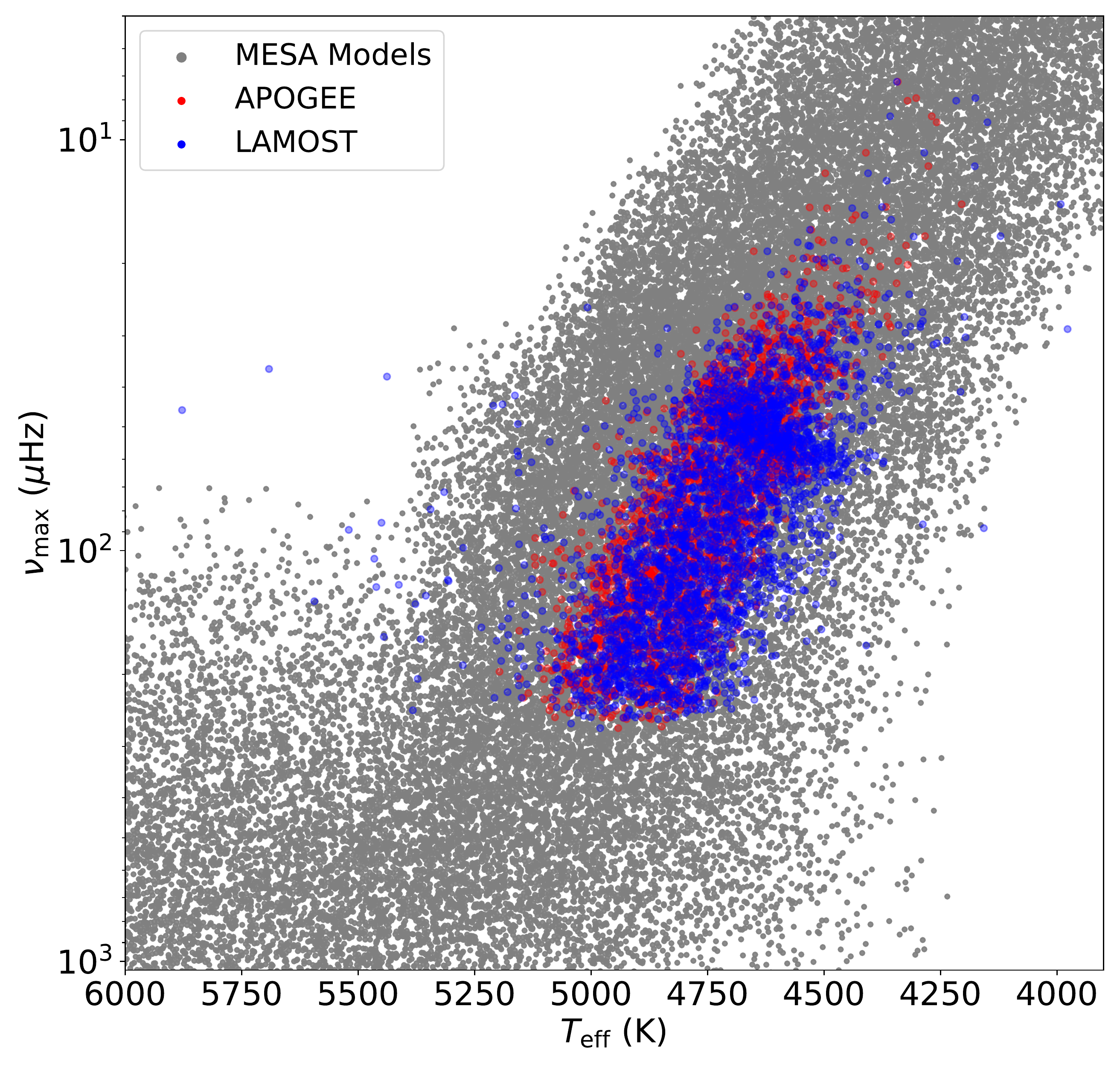}
\includegraphics[scale=0.32]{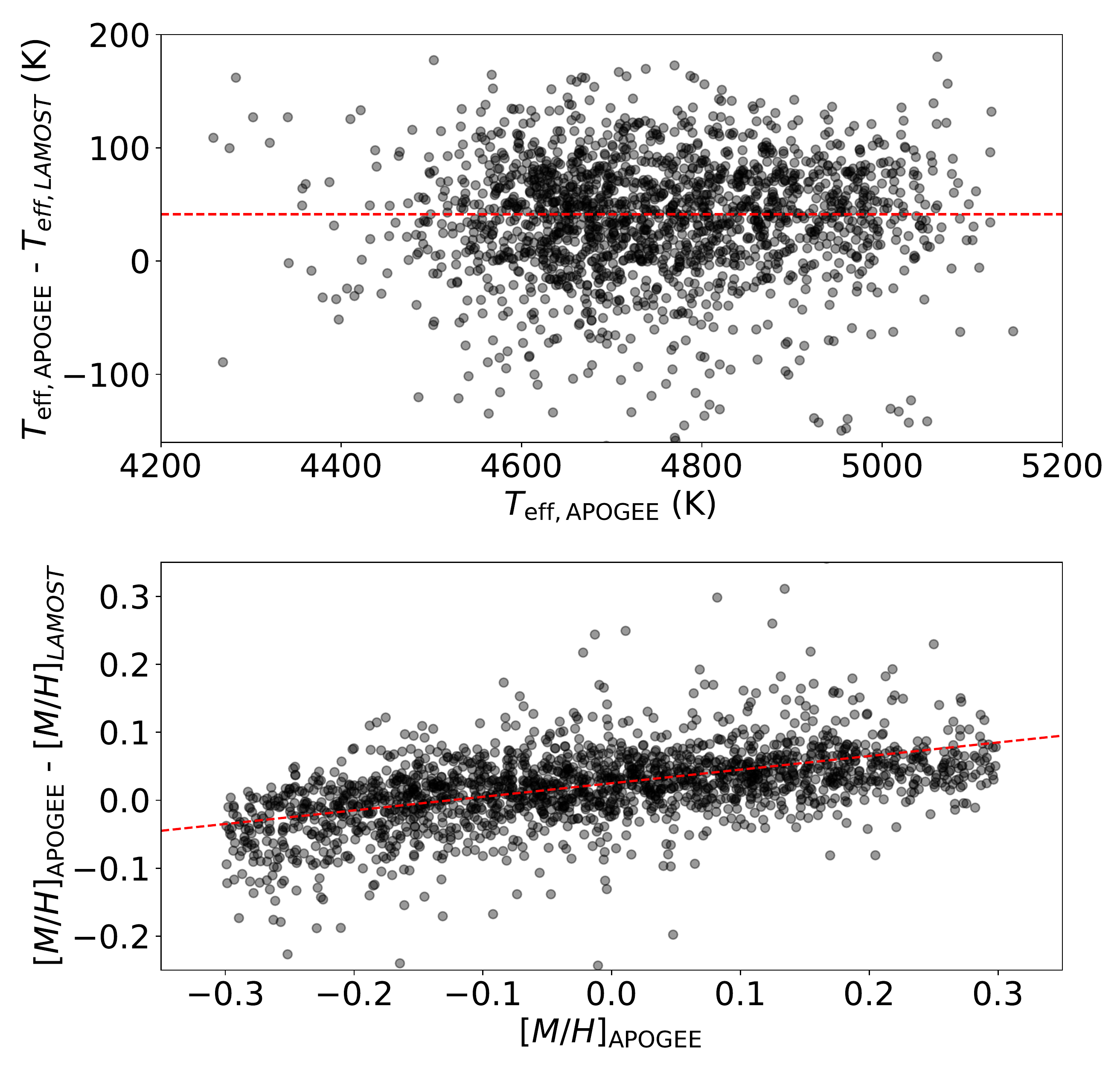}
\caption{Left: The \kepler{} RGB stars on the $T_{\rm eff}$--$\nu_{\rm max}$ diagram. Grey dots are 50,000 \textsc{MESA} models randomly selected from our grid. Blue and red dots represent the 2,626 $\apogee$ and 3,326 $\lamost$ tragets. Observed $\nu_{\rm max}$ values are from \citet{yuj++2018-16000-rg}.
Right: Differences in effective temperature (upper) and metallicity (bottom) of the 1,825 common stars. Red dashes indicate the systematic offsets, which is constant ($T_{\rm eff, APOGEE}$ = $T_{\rm eff, LAMOST}$ + 41 K) for effective temperature and a linear function ([M/H]$_{\rm APOGEE}$ = 0.20[M/H]$_{\rm LAMOST}$ + 0.023 dex) for metallicty.
\label{fig:hrd}}
\end{figure}

We started with \kepler{} RGB stars classified by \citet{2017MNRAS.469.4578H}, who used a machine-learning tool to distinguish between RGB stars and core-helium-burning stars, based primarily on the gravity-mode spacings \citep{bedding++2011-distinguish-rc-rgb,mosser++2012-core-and-evolution-rg-mixed-modes}. According to \citet{2017MNRAS.469.4578H}, their classification achieved an accuracy of up to 99\%.
These stars were studied by \citet{yuj++2018-16000-rg}, who measured their global seismic parameters (\Dnu{} and \numax{}) using the SYD pipeline \citep{huber++2009-syd-pipeline}.
We also included six oscillating red giants in EB systems, which were previously studied by \citet{2018MNRAS.475..981L}, as the benchmark stars in this study.
We cross-matched \kepler{} RGB sample with spectroscopically measured atmospheric parameters (\teff{} and [M/H]) from two large-scale spectroscopic surveys: \apogee{} DR16 \citep{ahunmada++2020-apogee-dr16} and \lamost{} DR5 \citep{xiang++2019-lamost-dr5}. 
We only selected stars within a metallicity range of [M/H] = -0.3--+0.3 to fall within the parameter range of our model grid (-0.5--+0.5).   
Note that the [M/H] values are not given in \lamost{} DR5, and we calculated them from [Fe/H] and [$\alpha$/Fe] values using the following formula, given by \citet{2005essp.book.....S}:  
\begin{equation}\label{eq:mh}
    {\rm [M/H]} = {\rm [Fe/H]} + \log(10^{\rm [\alpha/Fe]}0.694 + 0.396).
\end{equation}
The cross-match resulted in 4,127 \kepler{} RGB stars, comprising 2,626 \apogee{} and 3,326 \lamost{} targets, with 1,825 common sources. We show the sample on the \teff{}-\numax{} diagram in Figure \ref{fig:hrd}. The sample covers the lower to the upper RGB, with a \numax{} range from 10 to 270 $\mu$Hz. 
We also inspected the systematic differences in effective temperatures and metallicities between the two surveys. As shown on the right of Figure \ref{fig:hrd}, there is an approximately uniform offset in \teff{}, which can be described as $T_{\rm eff, APOGEE}$ = $T_{\rm eff, LAMOST}$ + 41K. The systematic offset in the [M/H] values follows a linear function. We fitted the data and obtained the relation as [M/H]$_{\rm APOGEE}$ = 0.20[M/H]$_{\rm LAMOST}$ + 0.023 dex. We show in Section \ref{sec:scaling} that this systematic offset has an impact on the correction of \numax{} scaling relation.           

% Gaia
% We also determine stellar luminosities based on trigonometric parallaxes with Gaia EDR3 \citep{gaia-2016-mission,gaia-2020-edr3}. We correct zero-point offsets in the parallaxes using a model provided by \citet{lindegren-2020-gaia-edr3-parallax-zpt}, and we multiple the uncertainties of the parallaxes by 1.3 to take into account of an underestimation of the uncertainties for brighter stars \citep{elbadry-2021-binary-gaia-dr3,zinn-2021-gaia-dr3-plx-seismology,maz-apellaniz-2021-gaia-dr3-plx-cluster}. Then we combine the parallaxes and 2MASS K-band magnitudes to determine the luminosities, using the ``direct'' method implemented in software \textsc{ISOCLASSIFY} \citep{huber++2017-seismic-radii-gaia,berger-2020-gaia-kepler-1-stars}.

% Fig.~\ref{fig:hrd} shows the H--R diagrams for the sample we studied in this work. The position of RGB bump is clear in both the \Dnu{}--\teff{} plane and the $L$--\teff{} plane, indicating the excellent precision of those measured parameters.

\subsection{Measurement of Radial Oscillation Mode Frequencies}
\begin{figure}
\centering
\includegraphics[width=0.6\textwidth]{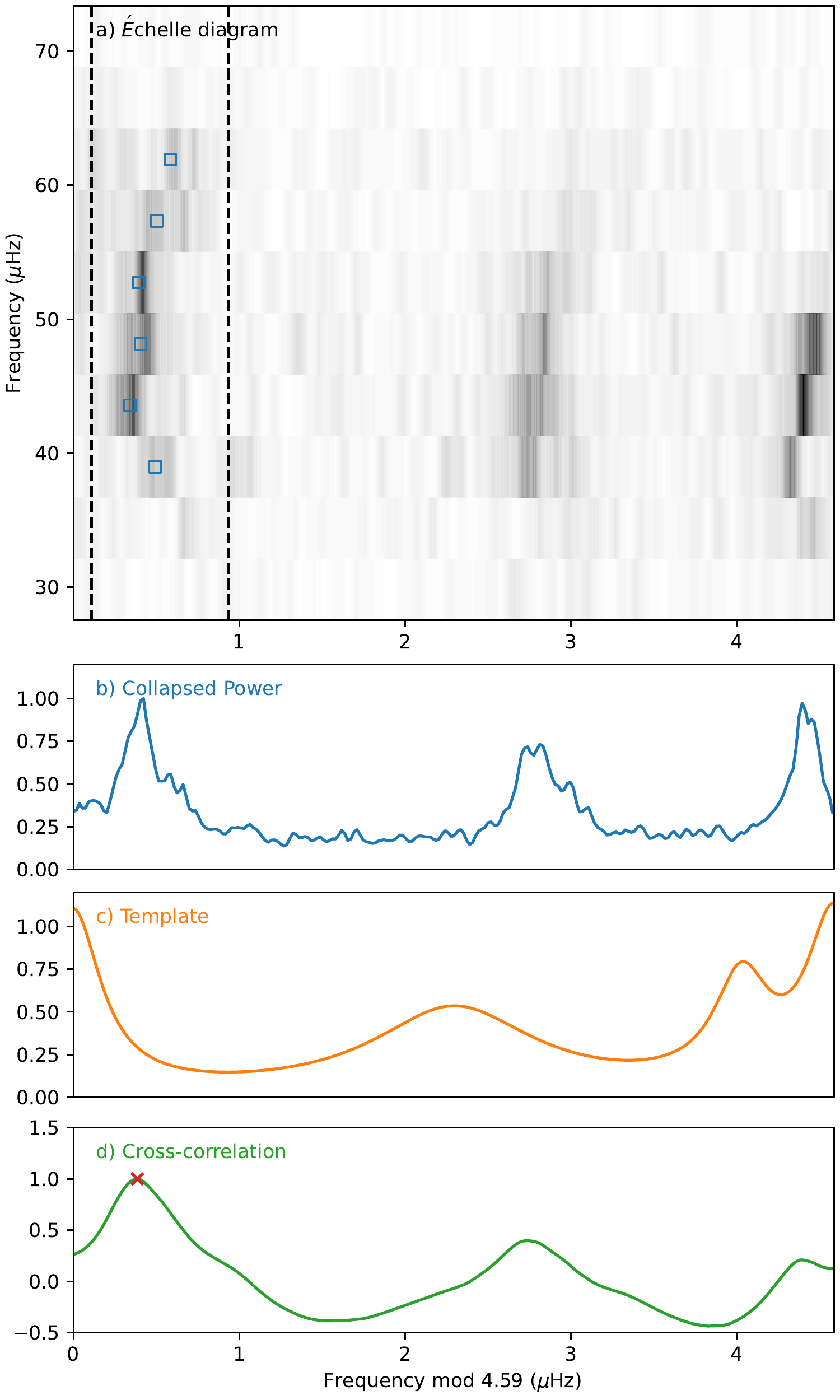}
\caption{Measurement of radial modes for KIC 8410637. a) The power spectrum in \'{e}chelle format. b) The collapsed power spectrum. c) The template spectrum. d) The cross-correlation spectrum between b) and c). The red cross shows the maximum in the cross-correlation spectrum, which gives the value of $\epsilon$ and identifies the ridge of radial modes. We used this to define the frequency ranges for the radial modes (shown as black dashed lines) and then fitted the strongest peaks within these ranges. The fitted frequencies are marked by squares. }
\label{fig:pipeline}
\end{figure}

% sample selection
We first identified the radial modes for the aforementioned stars. 
The asymptotic relation for p modes can be used to predict the mode frequencies on the basis that modes of the same $l$-degree and sequential $n$ are separated by about \Dnu{} \citep{tassoul-1980-asymptotic-relation,gough-1986-aymptotic-relation}:
\begin{equation}
	\nu(n,l) = \Delta\nu \left(n+\frac{l}{2}+\epsilon \right)  - \delta\nu_{0l},
\end{equation}
where $\epsilon$ defines an offset and $\delta\nu_{0l}$ is the small spacing between modes of different $l$-degrees.
We made use of this regularity to determine approximate locations of the radial modes, which means determining~$\epsilon$ for each star. We used the background-corrected power spectra produced by \citet{yuj++2018-16000-rg} and selected data around \numax\ in a range of 5 times the FWHM (full width at half maximum) of the power excess.
We then obtained the so-called collapsed power spectrum where the power was added from consecutive slices with width of \Dnu. This collapsed power spectrum was cross-correlated with a template comprising three peaks, constructed according to the asymptotic equation, corresponding to the ridges of $l=0$, 1 and~2 modes. 
We adopted a value for \dnu{0l} that was a fixed fraction of \Dnu{}, set by a fit to real stars \citep{huber++2010-800-rg-kepler}. The value of $\epsilon$ was then obtained by maximising the correlation coefficient. The locations of radial modes were restricted to be in the region $(\nu_{n,l=0}/\Delta\nu \ {\rm mod}\  1) \in [\epsilon-0.06, \epsilon+0.12 ] $. Fig.~\ref{fig:pipeline} shows this identification procedure for an example star, KIC 8410637. 

Having located the radial modes, we fitted Lorentzian profiles to these regions to extract the radial mode frequencies:
\begin{equation}
	M(\nu) = B(\nu)\cdot \eta^2(\nu) \cdot \sum_n L_{n}(\nu)  =B(\nu)\cdot \frac{\sin^2(\pi\nu\Delta t)}{(\pi\nu\Delta t)^2} \cdot \sum_n \frac{A_{n}^2/(\Gamma_{n})}{1+4(\nu-\nu_{n})^2/\Gamma_{n}^2},
\end{equation}
using the fitting method as described by \citet{liyg++2020-kepler-36-subgiants}.
We used an H1 (odds ratio) approach to evaluate the quality of each fit \citep{appourchaux++2012-freq-ms-sg-kepler,corsaro+2014-diamonds,davies++2016-bayesian-peakbagging-35-kepler-planethost,lund++2017-legacy-kepler-1}. We retained those modes with a Bayes factor larger than $3$, equivalent to strong detection \citep{kass&raftery}. { As an example, we listed model frequencies, uncertainties of KIC 8410637} in Table~\ref{tab:freqency_list}. This star is the red-giant component in a classified EB system. Its radial mode frequencies were previously measured by direct fitting \citep{2018MNRAS.475..981L}. The comparison shows that the model frequencies from our pipeline well agree with previous results and we find one extra mode at low frequency. We also found good consistencies in extracted modes for five other red-giants components studied by \citet{2018MNRAS.475..981L}. 

The number of identified radial modes is 6 on average, reaching up to 10 for low-luminosity RGB stars and decreasing with the evolution. For high-luminosity red giants with \numax{} below 20$\mu$Hz, most stars have only 3--4 modes. We note that a few of the stars may have incorrect results since they were returned from an automatic pipeline. They do not affect population studies because of their small number. However, to use frequencies for a particular star, they should carefully vetted. The source code for the mode measurement and lists of identified frequencies for the all stars are available on \url{https://github.com/litanda/SolarlikePeakbagging}.

%Fig.~\ref{fig:obs-modes} shows the extracted frequencies ranked as a function of \numax{}.
%\red{TODO: I suggest to have a section to summary our results like: 'We identify at least three individual radial modes in 5050 RGB stars out of the whole XXXX sample with a typical uncertainty of XXXX' explain 'Why we fail to identity mode from some stars whose $\Delta\nu$ can be extracted?' and give same general descriptions like: 'For star with the best data, we extract XXX orders of mode for low-luminosity red giants and XXX orders for high-luminosity ones.' Yaugang will add text for Fig. 3}

\begin{table}
\centering
\caption{Extracted radial modes of an example star KIC 10513837.}
\label{tab:freqency_list}

\begin{tabular}{ccc}
\hline
$\nu$ ($\mu$Hz) &  $\sigma_{\nu}$ ($\mu$Hz) & Ref.$^{a}$ \\
\hline
37.19&	0.02 &   -\\
41.62&	0.02 & $41.63\pm0.03$\\
46.28&	0.02 & $46.29\pm0.03$\\
50.85&	0.02 & $50.86\pm0.02$\\
55.55&	0.06 & $55.53\pm0.05$\\
60.22&	0.05 & $60.37\pm0.03$\\
\hline
% \multicolumn{3}{l}{$^{a}$The Bayes factor reflecting the}\\
% \multicolumn{3}{l}{significance of detection. } \\
\multicolumn{3}{l}{$^{a}$ Mode frequencies from \citet{2018MNRAS.475..981L}.}
\end{tabular}
\end{table}
% KIC 8410637
% 	PTfc	PTefc
% 0	37.190312	0.020955
% 1	41.622963	0.015278
% 2	46.276658	0.016648
% 3	50.850801	0.017140
% 4	55.547486	0.056961
% 5	60.216642	0.053079

\section{Theoretical Modeling}\label{sec:model}
\subsection{Model Grid and Input Physics} \label{subsec:grid}

We constructed a stellar model grid that ranges from 0.76 to 2.20$M_{\odot}$ with a mass step of 0.02$M_{\odot}$. The grid includes four independent model inputs: stellar mass ($M$), initial helium fraction ($Y_{\rm init}$), initial metallicity ([M/H]), and the mixing-length parameter ($\alpha_{\rm MLT}$). Ranges and grid steps of these four model inputs are given in Table \ref{tab:grid}. We started computation of each stellar evolutionary track from the Hayashi line and terminated on the upper RGB when \mbox{$\log g$ $\geq$ 1.5\,dex}. Note that we did not include models undergoing core helium burning.  

\begin{table}
\centering
\caption{Input Ranges and Grid Step of the Model Grid.}
\label{tab:grid}
\begin{tabular}{lccc} % four columns, alignment for each
\hline
Input Parameter & \multicolumn{2}{c}{Range} & Increment\\
& From & To & \\
\hline
$M / $M$_{\odot}$  & \phantom{$-$}0.76 & 2.20 &  0.02\\
$\rm{[Fe/H]}$ & $-0.5$\phantom{0} & 0.5\phantom{0} & 0.1\phantom{0}\\
$Y_{\rm init}$ & \phantom{$-$}0.24 & 0.32 & 0.02\\
$\alpha_{\rm{MLT}}$ & \phantom{$-$}1.7\phantom{0} & 2.5\phantom{0} & 0.2\phantom{0}\\
\hline
\end{tabular}
\end{table}

We used the stellar evolution code {\sc mesa} (Modules for Experiments in Stellar Astrophysics, version 12115, \citealt{2011ApJS..192....3P,2013ApJS..208....4P, 2015ApJS..220...15P,2018ApJS..234...34P}) and the oscillation code {\sc gyre} (version 5.1, \citealt{2013MNRAS.435.3406T}) to compute the model grid. We adopted the solar chemical mixture [$(Z/X)_{\odot}$ = 0.0181] provided by \citet{2009ARA&A..47..481A}. 
The initial chemical composition was calculated by:
\begin{equation}
\log (Z_{\rm{init}}/X_{\rm{init}}) = \log (Z/X)_{\odot} + \rm{[M/H]_{init}}.  \\
\end{equation}
We used the {\sc mesa} $\rho$--$T$ tables based on the 2005 update of the OPAL equation of state tables \citep{2002ApJ...576.1064R} and OPAL opacities supplemented by the low-temperature opacities from \citet{2005ApJ...623..585F}. The Eddington photosphere was used for the set of boundary conditions for modeling the atmosphere. The mixing-length theory of convection was applied, where $\alpha_{\rm MLT} = \ell_{\rm MLT}/H_{\rm p}$ is the mixing-length parameter. We considered convective overshooting at the core, the H-burning shell, and the envelope. The exponential scheme by \citet{2000A&A...360..952H} was applied and the diffusion coefficient in the overshoot region given as  
\begin{eqnarray}
{D_{\rm{OV}}} = {D_{\rm{conv,0}}}\exp \left( - \frac{{2(r-r_{0})}}{{(f_0 + f_{\rm{ov}}){H
_{p}}}}\right).
\end{eqnarray}
Here, $D_{\rm{conv,0}}$ is the diffusion coefficient from the mixing-length theory at a user-defined location near the Schwarzschild boundary. The switch from convection to overshooting was set to occur at $r_0$. To consider the step taken inside the convective region, $(f_0 + f_{\rm{ov}}){H_{\rm p}}$ was used, where $f_{\rm 0}$ equals 0.5$f_{\rm{ov}}$. 
Following \citet{2010ApJ...718.1378M}, the overshooting parameter is mass-dependent following a relation as $f_{\rm{ov}} = (0.13M - 0.098)/9.0$, and a fixed $f_{\rm{ov}}$ of 0.018 was adopted for models above $M = 2.0$\,M$_{\odot}$. For a smooth convective boundary, we also applied the {\sc mesa} predictive mixing scheme. The mass-loss rate on the red-giant branch with Reimers' law was set as $\eta = 0.2$, which is constrained by the seismic targets in old open clusters NGC\,6791 and NGC\,6819 \citep{2012MNRAS.419.2077M}. Atomic diffusion was only considered for models below 1.1M$_{\odot}$ at the main-sequence phase (it is turned off when central hydrogen fraction goes below 0.01). As discussed by \citet{2012A&A...541A..41M}, the mass value of 1.1M$_{\odot}$ is a proper switching point, above which the surface helium and heavy elements would be rapidly drained out of the envelope if atomic diffusion were applied. The switch of atomic diffusion causes a discontinuity in the surface metallicity as well as other global parameters during the main-sequence and early subgiant stages, but most of the effects are not significant at the RGB phase due to the redistribution of star interiors \citep{2010A&A...510A.104M,2017ApJ...840...99D}. 
However, element diffusion also changes the main-sequence lifetime and hence a discontinuity in star age is expected for the RGB models at 1.1 M$_{\odot}$ \citep[e.g. Fig. 11][]{2012A&A...541A..41M}. To investigate the systematic uncertainty, we compared stellar ages at given mean density on RGB for evolutionary tracks with and without atomic diffusion. The comparisons shows an 8\% difference for 1.1M$_{\odot}$ evolutionary tracks. The age difference is strongly mass-dependent, going down to 1\% at 1.2M$_{\odot}$ and is approximately zero at 1.3M$_{\odot}$. We hence expect a systematic uncertainty of a few per cent in inferred ages in the mass range of 1.1--1.2 M$_{\odot}$.
Compared with the typical uncertainty (16\%), our estimated ages for stars about 1.1M$_{\odot}$ could be biased to some extent for the choice of atomic diffusion, but this systematic difference is negligible above 1.2 M$_{\odot}$.

\subsection{Fitting Method}\label{sec:fitting}
\begin{figure}
\centering
\includegraphics[scale=0.27]{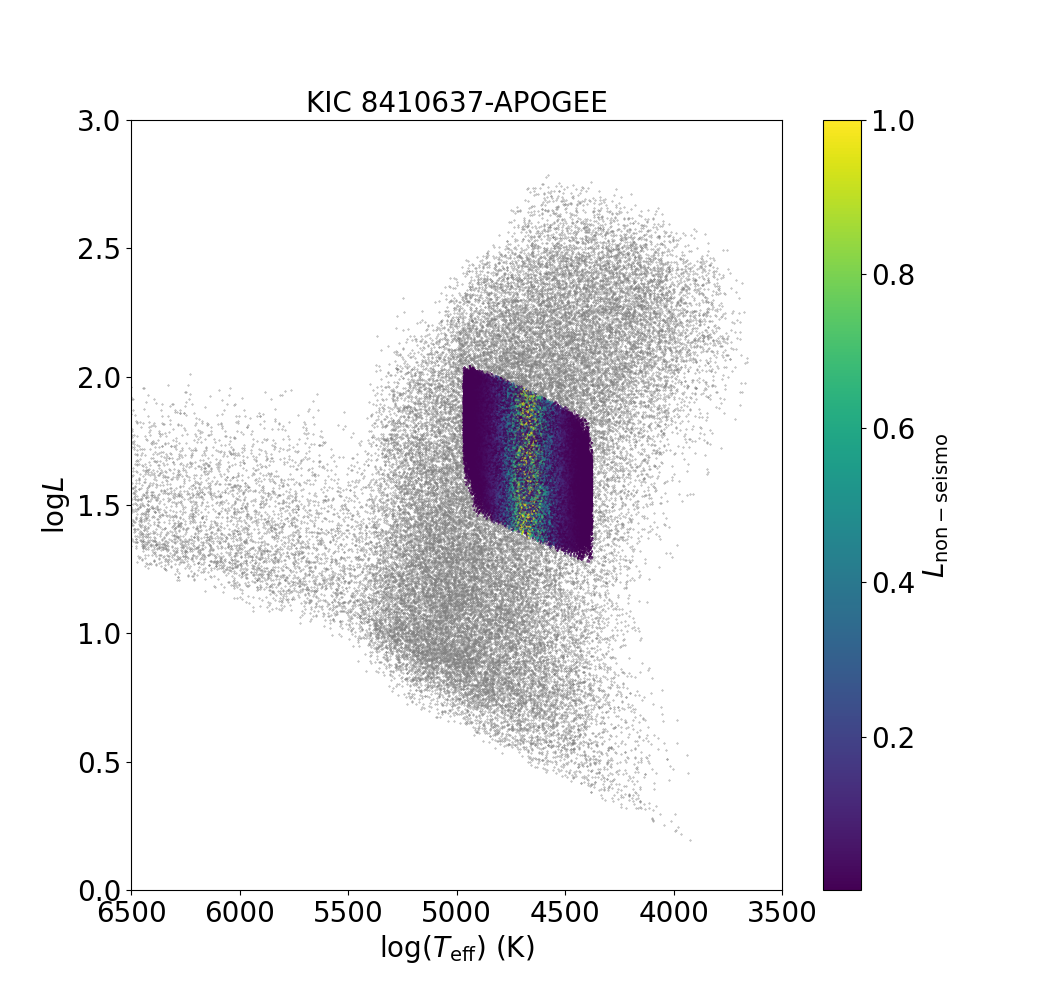}
\includegraphics[scale=0.27]{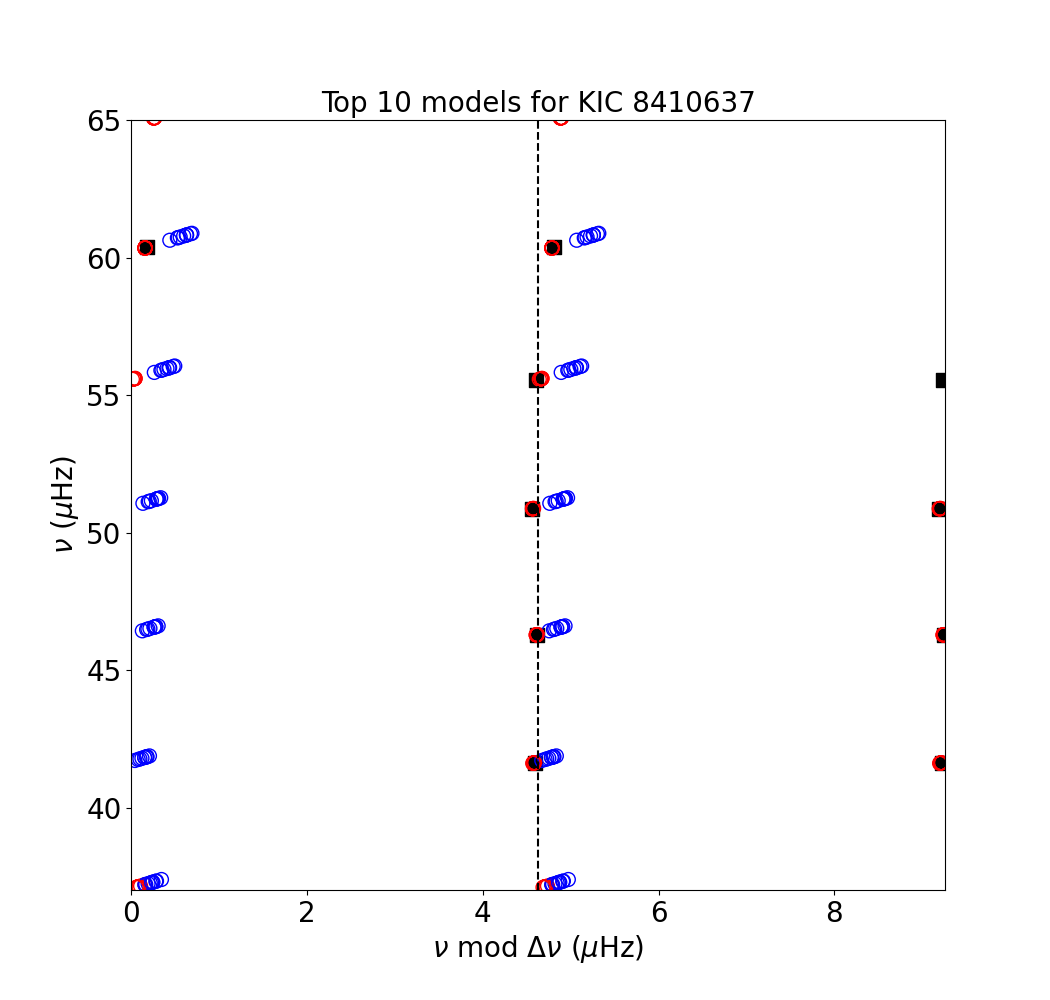}
\includegraphics[scale=0.27]{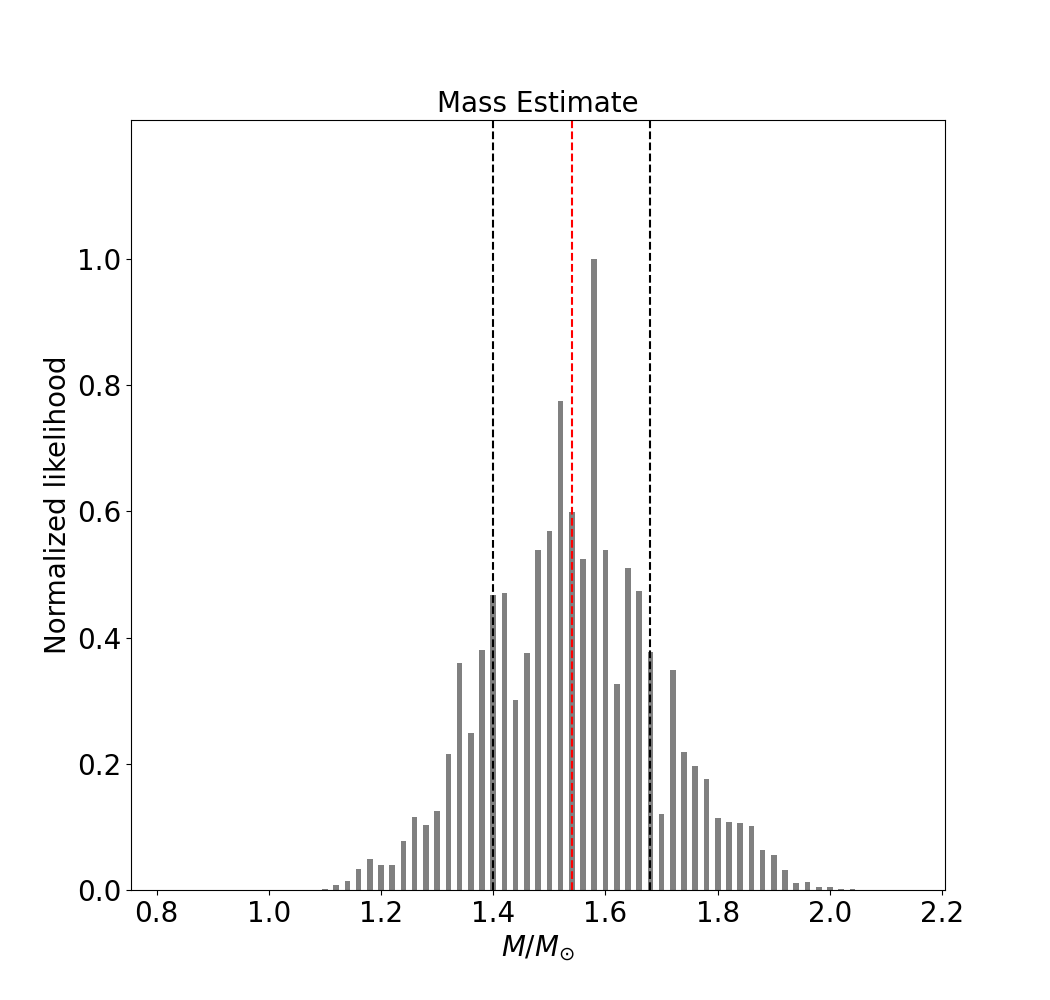}
\includegraphics[scale=0.27]{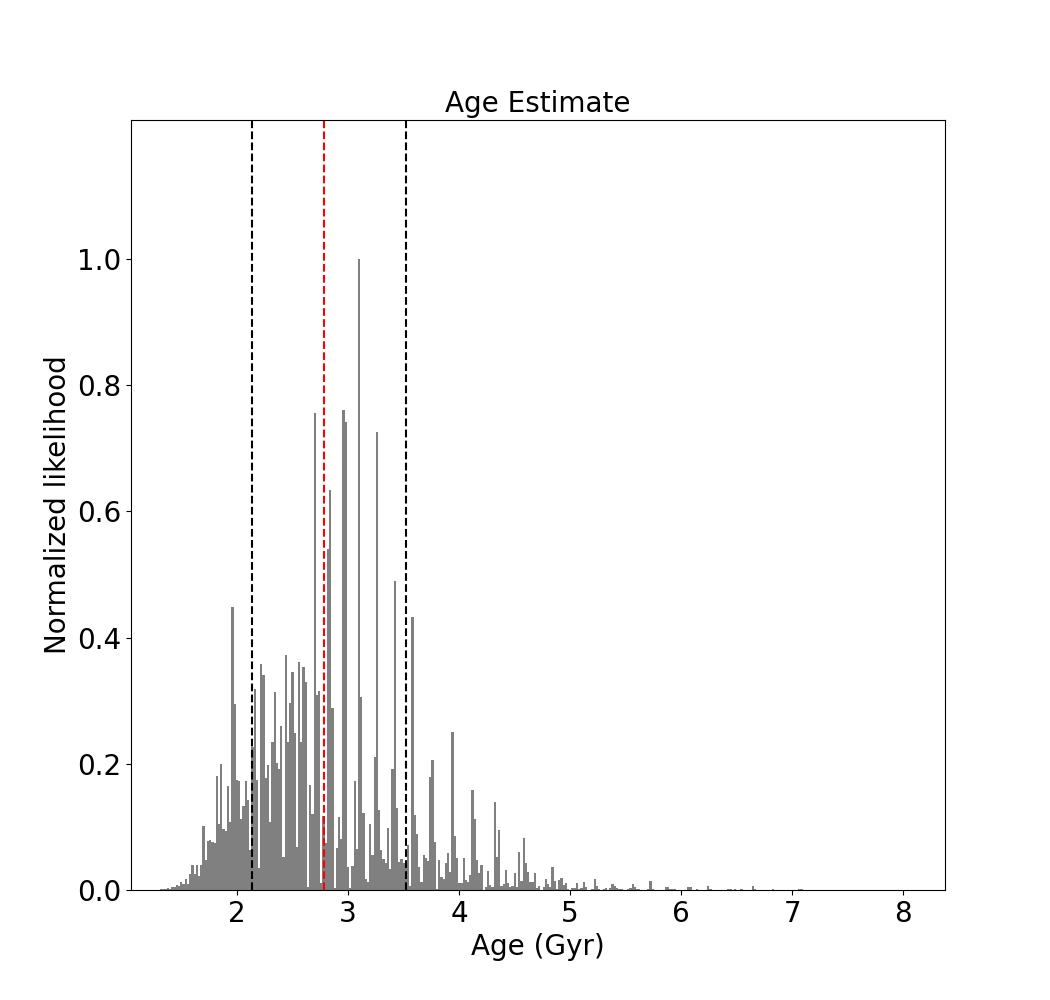}
\includegraphics[scale=0.27]{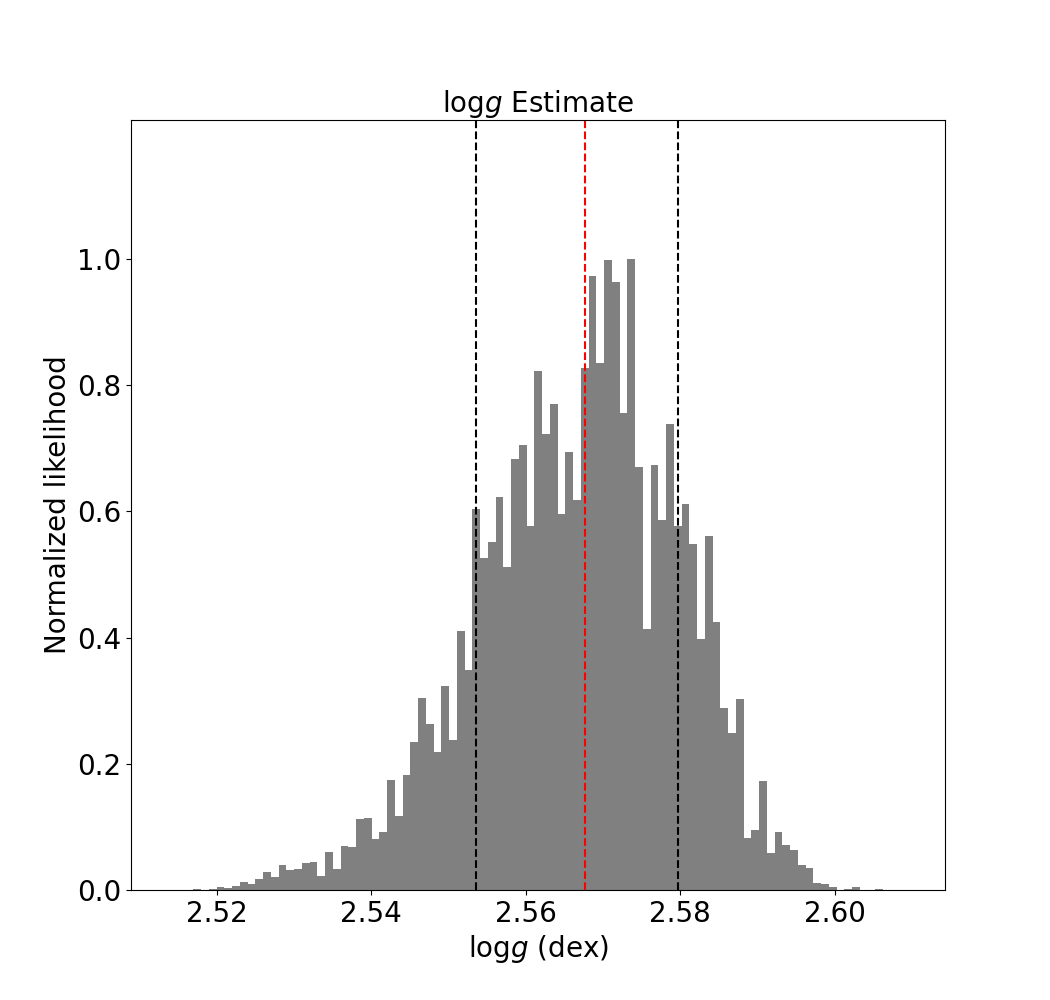}
\includegraphics[scale=0.27]{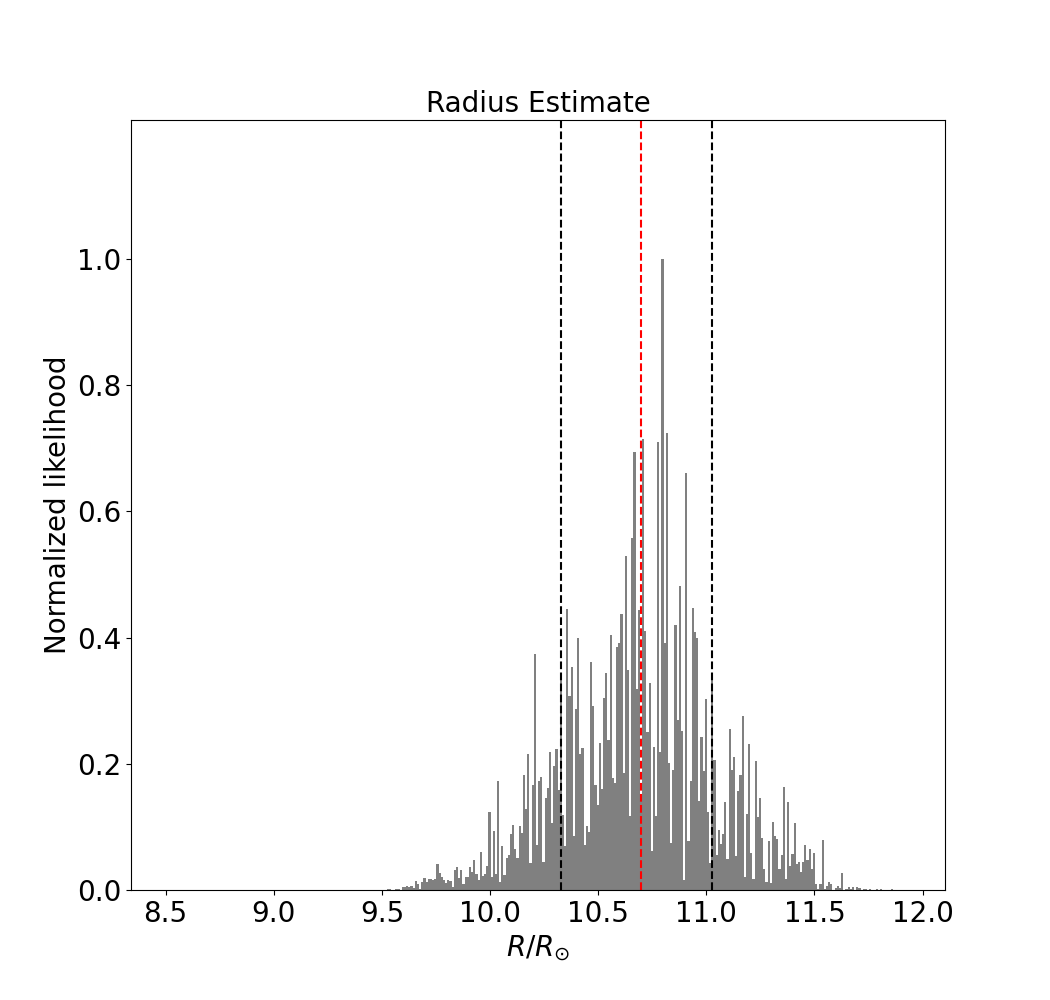}
\caption{The fitting procedure for the example star (KIC 8410637). Top left: Fitting models to spectroscopic constraints (\teff{} and [M/H]) in the range 0.8--1.2 $\Delta \nu$ and selecting those with $L_{\rm non-seismo} >0.0001$ for the subsequent seismic analysis. Grey dots represent the all models, colour-coded dots are selected models and colour indicates the non-seismic likelihood ($L_{\rm non-seismo}$). Top right: Fitting observed mode frequencies. We demonstrate the fits on the \'{e}chelle diagram. Filled black squares are observed mode frequencies, open blue and red circles represent theoretical mode frequencies before and after the surface correction. The \'{e}chelle diagram is plotted twice to fully present the radial mode pattern, and the blue dashed line indicates the large separation. Middle and bottom rows: probability distributions for mass, age, surface gravity, and radius. The central red dashed line indicates the median of each distribution, and two black dashed lines mark the 16\% and the 84\% cumulative values (1-$\sigma$ uncertainty).
\label{fig:fitting}}
\end{figure}

We used the Maximum Likelihood Estimate (MLE) method for fitting the models to the observations. In Figure~\ref{fig:fitting}, we demonstrate our fitting procedure with the example star. We first searched for models in the range 0.8--1.2 times $\Delta \nu$. Note that the $\Delta\nu$ scaling relation was used in this step and the $\pm$20\% range was large enough to cover its systematic uncertainty. We fitted the observed $T_{\rm eff}$ and [M/H] to models and calculated the non-seismic likelihood ($L_{\rm non-seismo}$) with the following equation:
\begin{equation}\label{eq:likelihood1}
    L_{\rm non-seismo} = {\rm exp}\left[ - \frac{(T_{\rm eff,obs} - T_{\rm eff,mod})^2}{2 \sigma^2_{T_{\rm eff,obs}}} - \frac{([M/H]_{\rm obs} - [M/H]_{\rm mod})^2}{2 \sigma^2_{[M/H]_{\rm obs}}} \right],  \\
\end{equation}
where the subscripts `obs' and `mod' stand for the observed and model values. 
Note that, because our grid resolution for [M/H] is 0.1 dex, an uncertainty of $\pm$0.1 for [M/H] was used when a star's observed uncertainty is smaller than this value. We selected models with $L_{\rm non-seismo} >0.0001$ for the subsequent seismic analysis.  

We corrected the surface term in oscillation frequencies of each selected model using the two-term formula given by \citet{2014A&A...568A.123B}. While inspecting the fits of frequencies, we noticed that some bad models whose frequencies deviated strongly from the observations could be made to fit well after an implausibly large surface correction. To avoid this, we set up a threshold for the surface term when selecting models for the seismic fitting. Although the surface term of a certain model is unknown before fitting, we can learn from previous studies to roughly estimate its range. According to results on \kepler{} main-sequence and RGB stars \citep{2018MNRAS.479.4416C,2018MNRAS.475..981L,2021MNRAS.501.3162L}, relative surface corrections at \numax{} ($\delta\nu_{\rm corr}$(\numax)/\numax) for most stars are in a range from $-0.005$ to $-0.01$. Considering the potential systematic differences between theoretical models, we used a larger range as the threshold, namely 0 to 0.02. After correcting the surface term, we checked the relative surface corrections at \numax{} and only kept models fitting this threshold criterion.

We next fitted individual radial mode frequencies and calculated a seismic likelihood of each model for as 
\begin{equation}\label{eq:likelihood2}
    L_{\rm seismo} = \prod {\rm exp}\left[ - \frac{(\nu_{i \rm,obs} - \nu_{i \rm,mod})^2}{ 2 (\sigma^2_{\nu_{i \rm ,obs}} + \sigma^2_{\nu_{\rm sys}}) } \right]. \\
\end{equation}
The subscript $i$ refers to the mode frequencies.
As discussed by \citet{2020MNRAS.495.3431L}, the systematic uncertainty in model frequencies cannot be neglected in the detailed modeling because it could be comparable to or larger than the observed uncertainty. For this reason, the fit was done in two steps. 
In the first step, we only considered the observed frequency uncertainty ($\sigma_{\nu,obs}$) in the likelihood function and found the best-fitting model. We used the median value of frequency offsets between the observations and this best-fitting model as the systematic uncertainty ($\sigma_{\rm sys}$). In the second step, we fitted to individual mode frequencies again but included the $\sigma_{\rm sys}$ term in the likelihood function.
Multiplying the non-seismic and seismic likelihoods gave us the final likelihood as
%\begin{equation}
$L = L_{\rm non-seismo}\cdot L_{\rm seismo}$.
%\end{equation}
We used the marginal likelihood distribution to measure the 16\%, 50\% and 84\% cumulative values to determine global parameters (i.e., mass, age, radius, surface gravity, and luminosity) and their uncertainties. 

\section{Results}\label{sec:results}

\subsection{Five Red Giants in Eclipsing Binaries}

\begin{figure}
\centering
\includegraphics[scale=0.6]{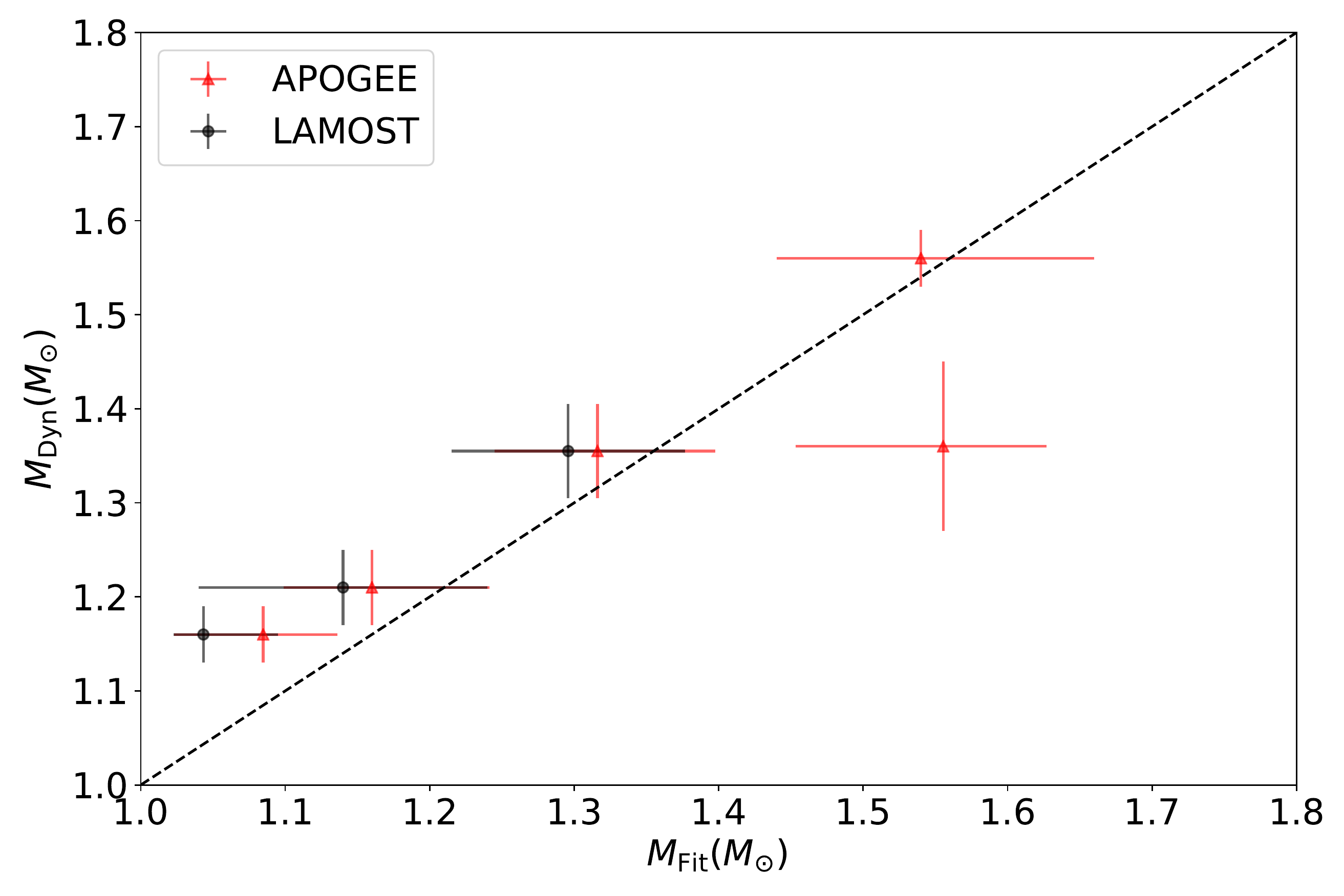}
\includegraphics[scale=0.6]{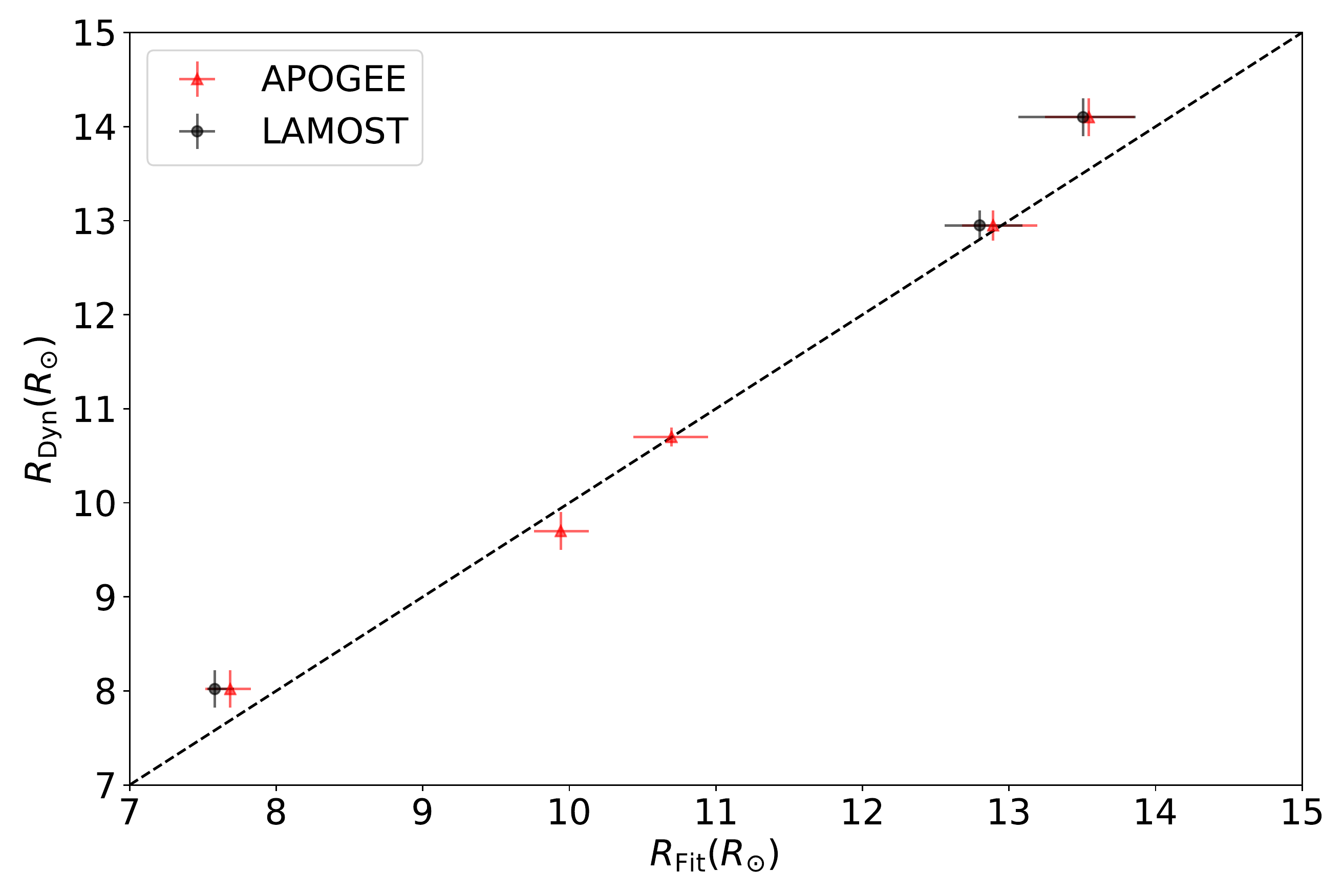}
\caption{Masses and radii determined by fitting radial modes versus those from dynamical modeling, for five red giants in EB systems. The black dashed lines show the 1:1 relation.
\label{fig:eb}}
\end{figure}

The accuracy of estimated masses and radii is critical because we use them to calibrate the scaling relations. We hence examined the accuracy of modeling solution with five red giants in EB systems, whose masses and radii have been previously determined with dynamical modeling.
%
% The first test we do with the five EBs is to check whether the {\it GAIA} luminosity gives useful constraint to models. The measurements of luminosity from {\it GAIA} DR3 present high precision, but the potential systematic uncertainty in {\it GAIA} parallax seen in the DR2 data \citep[e.g.][]{stassun2018evidence} is concerned. We model the five RGB components in EB systems. With the {\it GAIA}   luminosity, the uncertainty of inferred masses and radii are obviously improved, however, we find poor modeling solutions on two out of the five stars. We present details in Appendix \ref{app:A1}.
% Although results of EBs can not represent case for the whole {\it GAIA} sample given that the measurement of luminosity is relatively poor for binaries compared with those for single stars, the offsets on two stars make us concern about the potential systematic offset on a star-to-star bias.
% For this reason, we decide not use the {\it GAIA} luminosity as an observed constraint. Thus, our observed constraints includes the effective temperature, the metallicity, and radial oscillation modes.
%
%Here we examine our modeling-determined masses and radii.
We found \apogee{} spectra available for all five red giants and \lamost{} data for three of them. 
The comparison is demonstrated in Figure \ref{fig:eb}. We used the dynamical results determined by \citet{2013A&A...556A.138F}, \citet{2016ApJ...832..121G} and \citet{2018MNRAS.476.3729B}, and the average value was taken when there was more than one estimate for the same star.  
With the {\it APOGEE} spectra, our model-based masses and radii agree within $1\sigma$ for three out of the five stars, $1.5\sigma$ for KIC 9970396, and $2\sigma$ for KIC 4663623. The reason for the large offset on KIC 4663623 could be the very different metallicities adopted in our fitting (${\rm [M/H]} = 0.08$) and the dynamical modeling (${\rm [M/H]} = -0.13$).    
Estimates for the three stars with the \lamost{} data are slightly more discrepant but still consistent with dynamical determinations. 
In Table \ref{tab:ebs}, we summarise estimated masses and radii given by the detailed modeling, the scaling relations, and dynamical analysis. Compared with the scaling relation, the fits to radial mode frequencies significantly improve the accuracy of mass and radius estimates. 
%
% It infers that fits to radial mode frequencies is able to correct the large offsets in the scaling relations.
%The results indicate good accuracy of our modeling method. 
%and hence the model-based masses and radii could properly correct the scaling relations. 
% The offsets range from 1.2 to 6.5 per cent for the mass and 0 - 4 per cent for the radius. On average, the offsets are much less than the systematic uncertainty in scaling relation found by \citet{2016ApJ...832..121G} and \citet{2018MNRAS.476.3729B}, which is around +15\% for mass and +5\% for radius. This result indicates that our method can determine mass and radius with good accuracy.

\begin{table}
\centering
\caption{Masses and radii of five red giants in EB systems derived from our modeling, the scaling relations, and dynamical modeling.}
\label{tab:ebs}
\begin{tabular}{cccccccc}
\hline
KIC & $M_{\rm Fit}$ & $R_{\rm Fit}$&  $M_{\rm SR}$ & $R_{\rm SR}$ &  $M_{\rm Dyn}$ &  $R_{\rm Dyn}$ & Ref. \\
& (M$_{\odot}$)&  (R$_{\odot}$)& (M$_{\odot}$)&  (R$_{\odot}$)& (M$_{\odot}$)&  (R$_{\odot}$)& \\
\hline
4663623 &  $1.56^{+0.10}_{-0.12}$ &   $9.94^{+0.20}_{-0.25}$ & 1.87$\pm$0.13 &  10.78$\pm$0.28 &   1.36$\pm$0.09 & 9.7$\pm$0.2 & G16 \\
\hline
7037405 &  $1.16^{+0.08}_{-0.08}$ &  $13.55^{+0.37}_{-0.31}$ & $1.38\pm0.10$ &  $14.76\pm0.41$ &   $1.25\pm0.04$ &     $14.1\pm0.2$ & G16 \\
&&&&&$1.170\pm0.020$ &     $14.000\pm0.093$&B18\\
\hline
8410637 &  $1.54^{+0.14}_{-0.14}$ &  $10.70^{+0.33}_{-0.37}$ & $1.78\pm0.19$ & $11.45\pm 0.56$ &  $1.56\pm0.03$ &     $10.7\pm0.1$ & F13 \\
\hline
9540226 &  $1.32^{+0.08}_{-0.08}$ &  $12.89^{+0.25}_{-0.28}$ & $1.58\pm0.11$ & $14.05\pm0.37$ &   $1.33\pm0.05$ & $12.8\pm0.1$ & G16\\
&&&&&$1.378\pm0.038$ & $13.060\pm0.160$&B18 \\
\hline
9970396 &  $1.08^{+0.06}_{-0.06}$ & $7.69^{+0.15}_{-0.16}$ & $1.43\pm0.08$ & $8.68\pm0.17$ &   $1.14\pm0.03$ &      $8.0\pm0.2$ & G16\\
&&&&&$1.178\pm0.015$ & $8.035\pm0.074$ & B18 \\
\hline
\end{tabular}
References: F13 -- \citet{2013A&A...556A.138F}, G16 -- \citet{2016ApJ...832..121G}, B18 -- \citet{2018MNRAS.476.3729B}
\end{table}

\subsection{Estimated Stellar Parameters}

We used the fitting method described in Section \ref{sec:fitting} to estimate stellar parameters (mass, age, surface gravity, and radius) for the 4,127 $\kepler$ RGB stars. 
We first removed three types of stars from final results: i) stars with three or fewer identified radial modes, because the fits will largely affected by the uncertainty in surface correction; ii) stars with estimated masses below 0.85 or above 2.10 M$_{\odot}$, because those values are close to the edge of the grid; iii) stars with fewer than 100 models for the seismic analysis, because their observed constraints are either close to the edge or outside the parameter ranges of the model grid.
We then examined age determinations to identify those targets with unrealistically large estimates. These targets may bias the calibration for the scaling relation, given that age correlates strongly with mass for RGB stars. 
We found 61 out of 5,350 estimated ages that are larger than the universe age (13.8 Gyr) by more than 2-$\sigma$. 
We inspected those targets and found some of them do not have reliable \teff{} measurements and others could be core-helium-burning stars according to their power spectra. The fraction ($\sim 1.1\%$) of poor age determinations generally agrees with the accuracy rate of classifications reported by \citet{2017MNRAS.469.4578H}.
{ We then removed the targets with poor age estimates from the sample.}
We finally obtained 5,289 modeling solutions for 3,642 stars, including 2,498 \apogee{} and 2,791 \lamost{} targets with 1,647 common sources. We list observed constraints and estimated parameters in Table \ref{tab:est} (the full table is included in the supplement).

\begin{table}
\centering
\caption{Observed constraints and model-inferred stellar parameters (full table available on-line).}
\label{tab:est}
\begin{tabular}{cccccc|ccccc}
& \multicolumn{5}{c}{Observed Constraints} & \multicolumn{4}{c}{Estimates} \\
\hline
KIC & Source & $T_{\rm eff}$ & [M/H]  &$\Delta\nu$ & $\nu_{\rm max}$ &  $M$ & $\tau$ & $R$ &  $\log g$\\
 & &(K) & (dex) & ($\mu$Hz)& ($\mu$Hz)& ({\rm $M_{\odot}$})&  (Gyr)& ({\rm $R_{\odot}$})& (dex)\\
\hline
8161920 & \apogee{} & 4551 & -0.046 &  3.65 &  31.9 & $1.14^{+0.08}_{-0.08}$ & $7.7^{+3.3}_{-1.5}$ & $11.26^{+0.22}_{-0.33}$ & $2.393^{+0.007}_{-0.011}$\\
 5358184 &  \apogee{} &  4632 &  0.104 &  6.23 &  62.4 & $1.14^{+0.02}_{-0.03}$ & $9.7^{+1.2}_{-1.8}$ & $7.87^{+0.08}_{-0.09}$ & $2.701^{+0.004}_{-0.004}$\\
10793872 &  \apogee{} &  4780 & -0.229 &  14.86 &  181.7  & $0.94^{+0.03}_{-0.04}$ & $14.6^{+2.9}_{-1.0}$ & $4.19^{+0.04}_{-0.09}$ & $3.17^{+0.003}_{-0.008}$\\
8815290 &  \apogee{} &  4821 & -0.112 &  4.46 &  45.5 & $1.72^{+0.10}_{-0.16}$ & $1.9^{+0.5}_{-0.3}$ & $11.38^{+0.25}_{-0.37}$ & $2.56^{+0.008}_{-0.012}$\\
7749249 &  \apogee{} &  4703 &  0.099 &  6.66 &  71.8  & $1.22^{+0.08}_{-0.04}$ & $5.1^{+1.2}_{-0.7}$ & $7.72^{+0.17}_{-0.11}$ & $2.751^{+0.007}_{-0.007}$\\
\hline
 8161920 &  \lamost{} &  4560 & 0.249 &  3.65 &  31.9 & $1.14^{+0.08}_{-0.04}$ & $8.6^{+2.5}_{-1.5}$ & $11.25^{+0.23}_{-0.19}$ & $2.394^{+0.006}_{-0.007}$\\
5358184 &  \lamost{} &  4573 & -0.018 &  6.23 &  62.4 & $1.14^{+0.02}_{-0.02}$ & $9.08^{+1.9}_{-0.9}$ & $7.87^{+0.06}_{-0.05}$ & $2.702^{+0.001}_{-0.003}$\\
10793872 &  \lamost{} &  4925 &  0.110 &  14.86 &  181.7 & $0.94^{+0.03}_{-0.04}$ & $16.2^{+1.4}_{-2.6}$ & $4.16^{+0.07}_{-0.06}$ & $2.17^{+0.005}_{-0.006}$\\
8815290 &  \lamost{} &  4749 & -0.195 &  4.46 &  45.5 & $1.68^{+0.12}_{-0.14}$ & $2.4^{+0.4}_{-0.5}$ & $11.26^{+0.31}_{-0.33}$ & $2.558^{+0.010}_{-0.009}$\\
7749249 & \lamost{} &  4681 &  0.117 &  6.66 &  71.8 & $1.24^{+0.08}_{-0.06}$ & $4.8^{+1.2}_{-0.5}$ & $7.76^{+0.17}_{-0.14}$ & $2.752^{+0.007}_{-0.007}$\\
\hline
\end{tabular}
\end{table}

\begin{figure}
\includegraphics[scale=0.75]{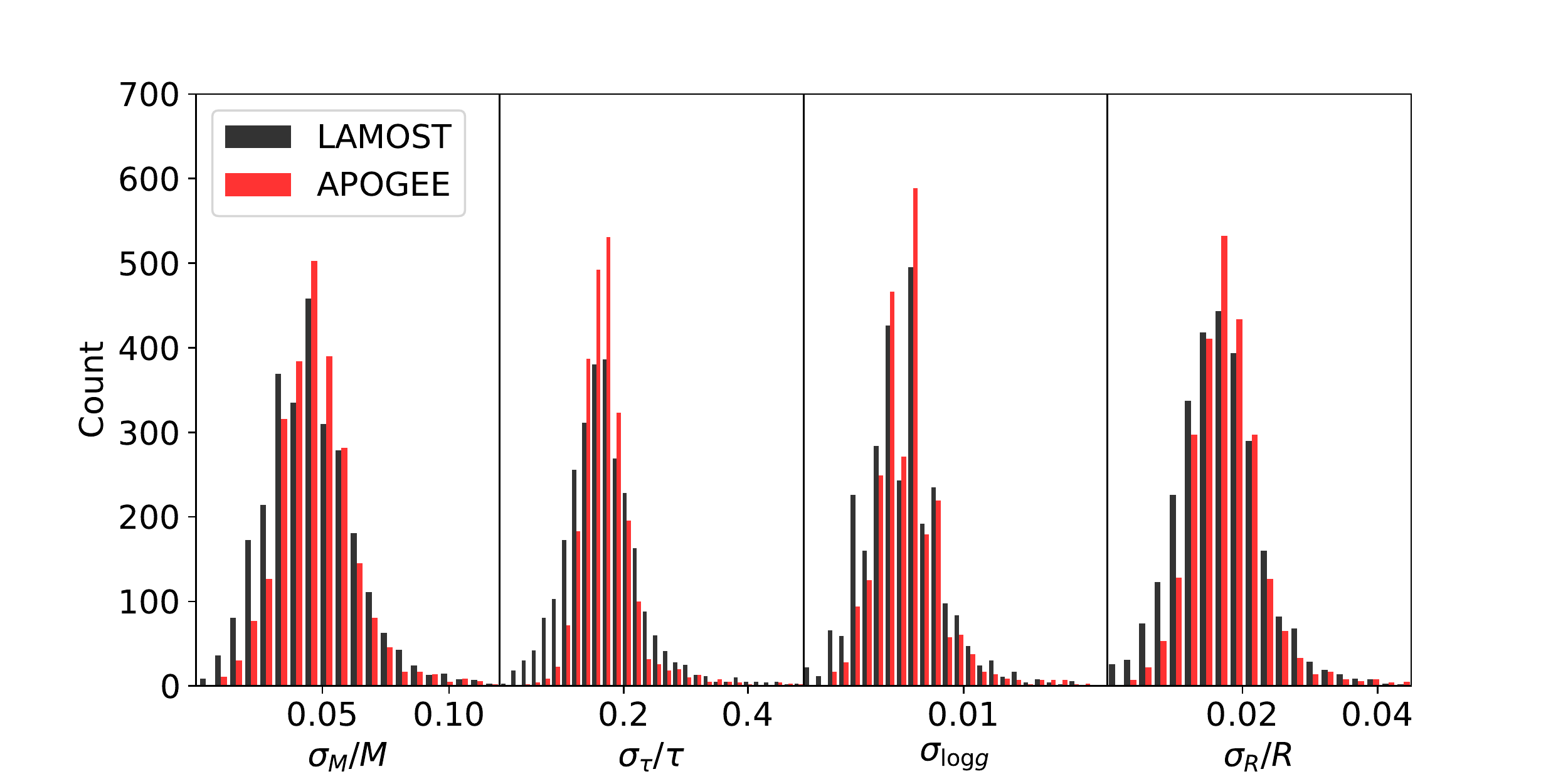}
\caption{Fractional uncertainties of the estimated parameters for stars with $\lamost$ and $\apogee$ spectroscopic constraints.
\label{fig:precision}}
\end{figure}

To inspect the precision of inferred parameters, we show histograms of the uncertainties in Figure \ref{fig:precision}. We obtained similar average precision for the \lamost{} and \apogee{} samples, but the \lamost{} sample spreads over a larger range. The typical precision (the median of the distribution) for both samples is 4.5\% for mass, 16\% for age, 0.006 dex for $\log g$, and 1.7\% for radius. 

\begin{figure}
\includegraphics[scale=0.55]{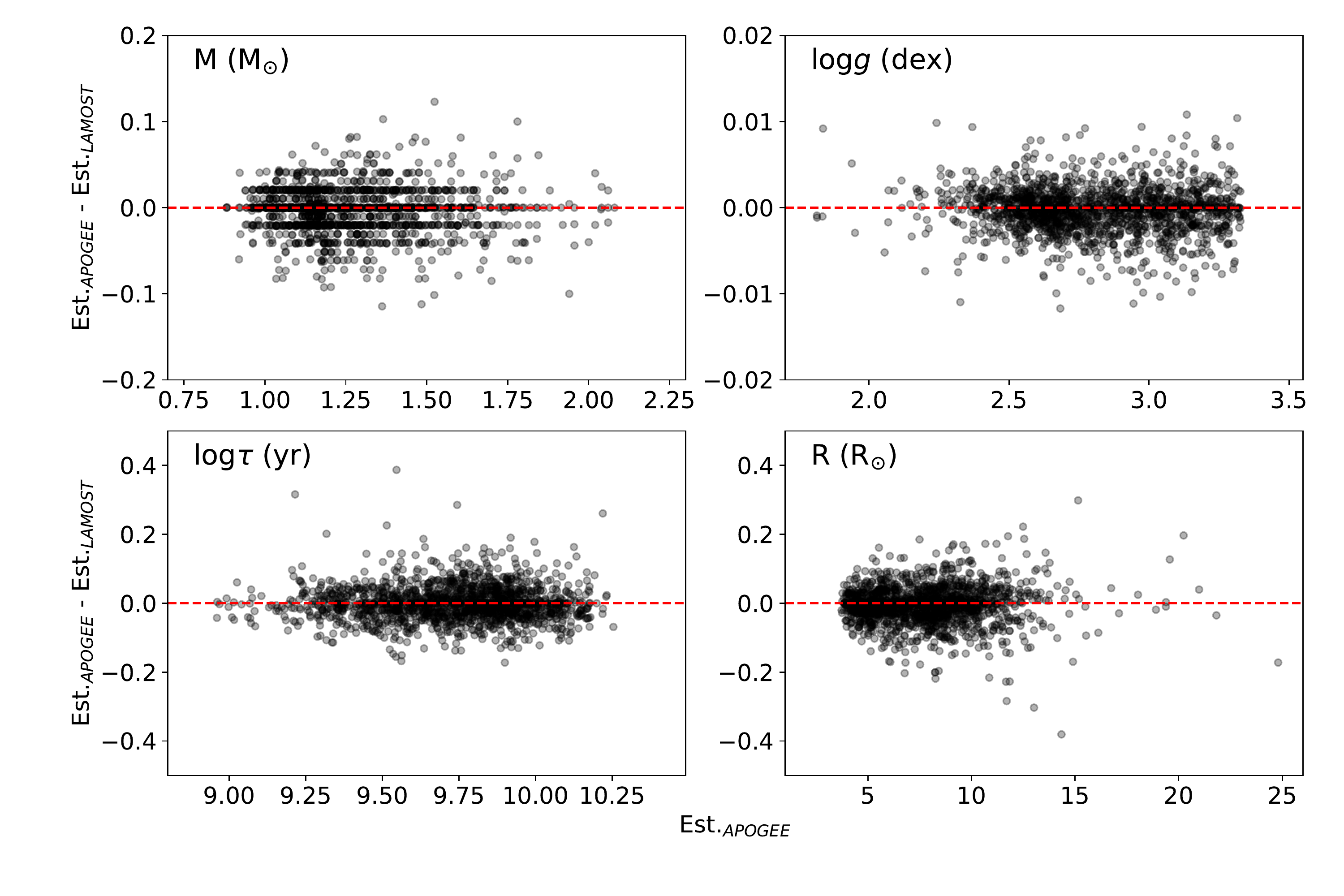}
\caption{Comparisons between estimated parameters with $\lamost$ and $\apogee$ spectroscopic constraints. (In the top left panel, some data points present stripe-like features. These stars are low-luminous red giants, whose mass loss is negligible, and hence their estimated masses are about the same as the initial masses, which are uniformly spaced. Other high-luminous stars have signficant mass loss, and hence their inferred masses distribute more randomly.) 
\label{fig:comparison}}
\end{figure}

We compared inferred values determined with the \apogee{} and the \lamost{} spectra to investigate any systematic differences caused by the spectroscopy. In Figure \ref{fig:comparison}, we illustrate comparisons for the 1,655 common targets and find good consistency at the sample level: the average differences for the four parameter are all around zero, with standard deviations that are similar to the typical uncertainties. The large offsets on some stars are due to the very different effective temperatures given by the two surveys.   
The results indicate that our fits are mainly determined by seismic mode frequencies, while the effective temperature and metallicity are relatively weak constraints.  

\section{The scaling relations for red giants}\label{sec:scaling}

\begin{figure}
\includegraphics[scale=0.35]{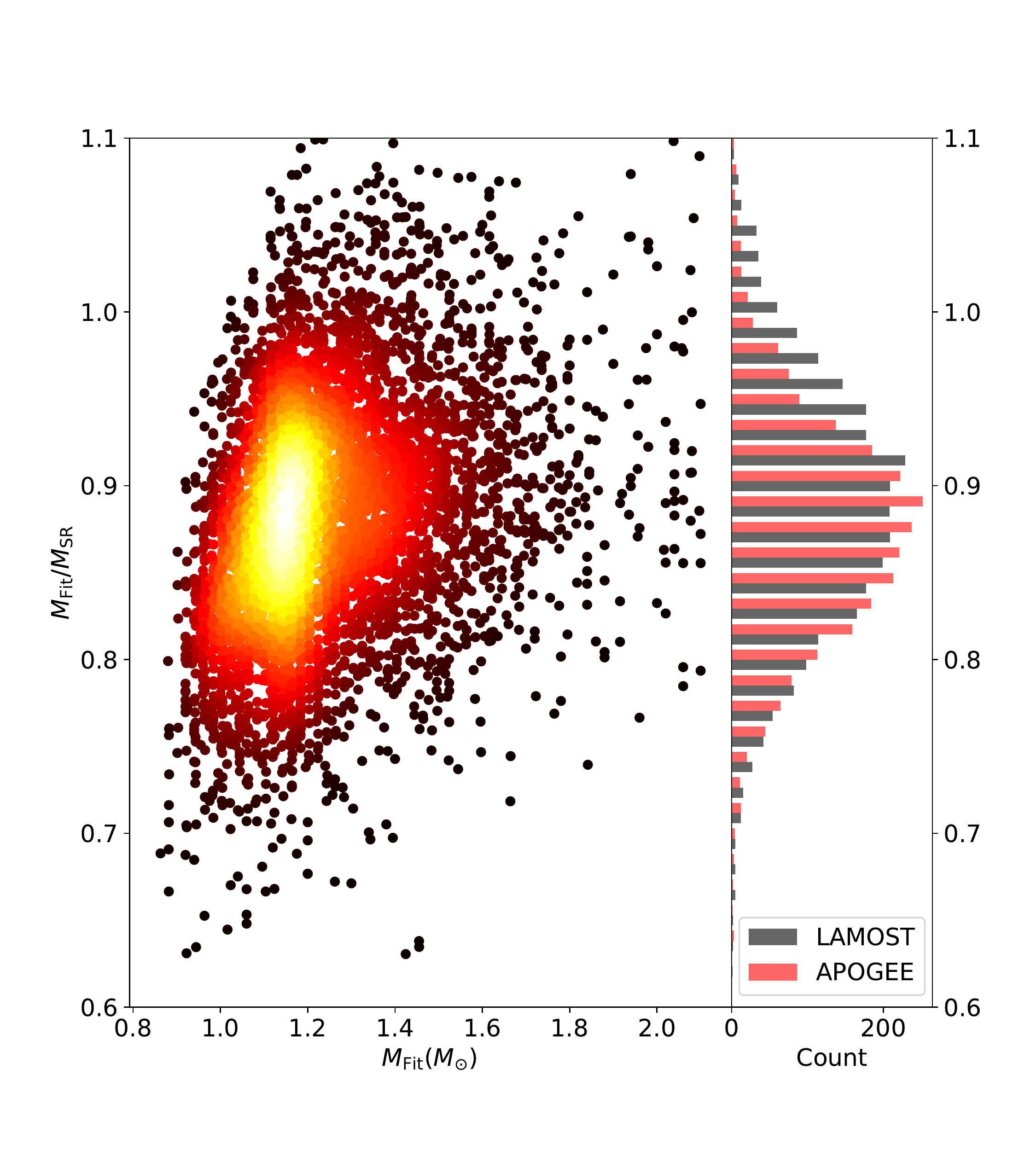}
\includegraphics[scale=0.35]{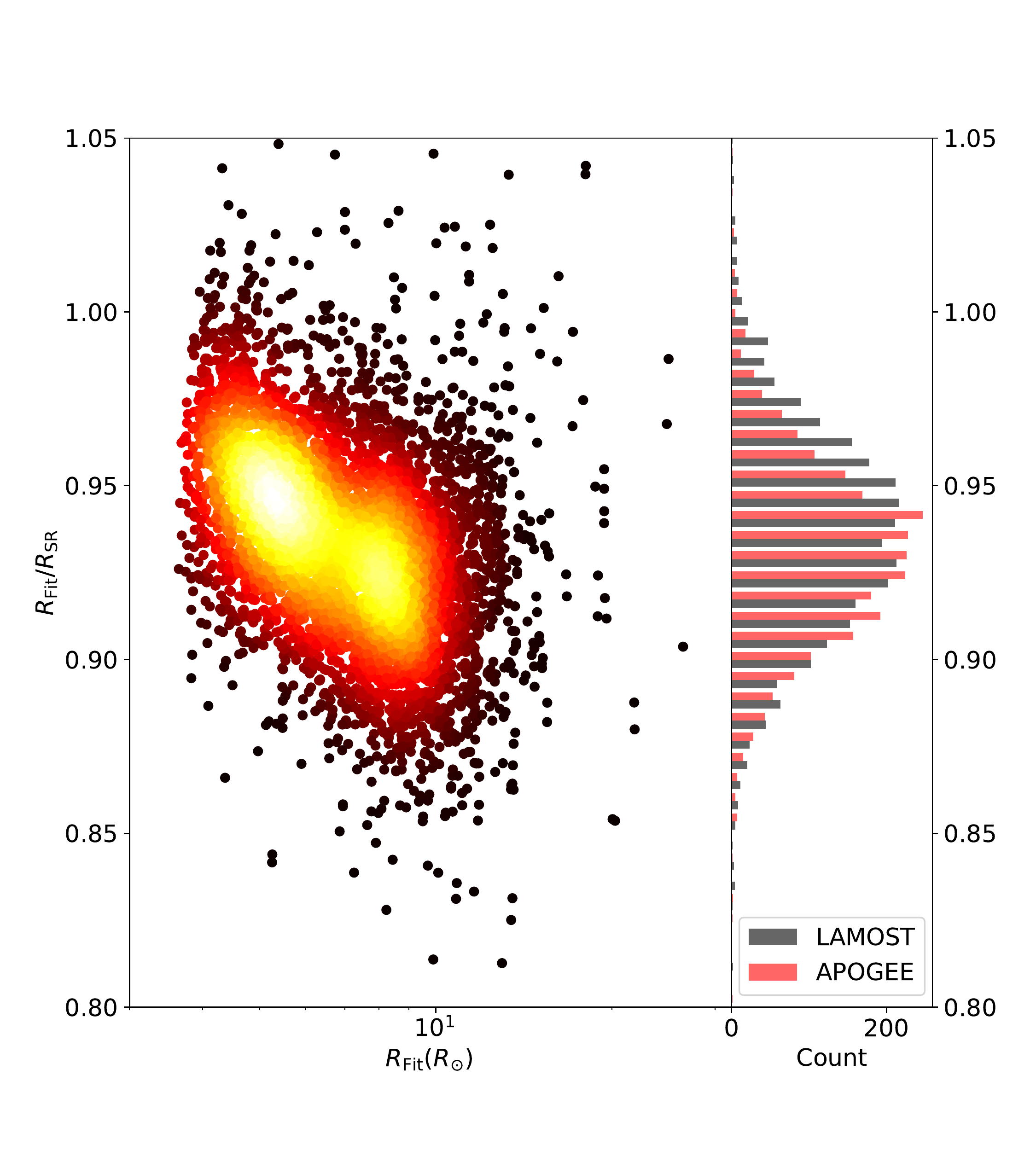}
\caption{Ratios between estimated masses and radii from our modeling and from the scaling relations. { Histograms show the stars with spectroscopic constraints from LAMOST and APOGEE.}
\label{fig:scaling_dif}}
\end{figure}

Our model-based masses and radii of the \kepler{} RGB star sample can allow testing of the scaling relations over a wide parameter range. We have shown that our estimates for stellar masses and radii are accurate, and hence they are good reference values for correcting the scaling relations. 
We start by inspecting the ratio between the modeling determinations and the scaling relations. As illustrated in Figure \ref{fig:scaling_dif}, we find that the scaling relations overestimate masses and radii for most of the stars. The density distributions give median factors of $\sim$85\% for the mass and $\sim$93\% for the radius. These values are consistent with those found for EBs \citep{2016ApJ...832..121G, 2018MNRAS.476.3729B, 2018MNRAS.478.4669T}. 
In Figure \ref{fig:scaling_dif}, we see that the ratio { weakly depends} on mass and this may correspond to some systematic offset caused by the effective temperature or the metallicity. On the other hand, the ratio clearly varies with stellar radius, indicating the discrepancy depends on the evolutionary phase.

We wish to correct the scaling relations by combining the modeling solution and observations. Rather than correcting Eq. \ref{eq:sc1} and \ref{eq:sc2} directly, we prefer to work with the $\nu_{\rm max}$  
%($\nu_{\rm max}$ as a function of $T_{\rm eff}$ and $g$) 
and the 
% $\Delta \nu$-$\rho$ 
\Dnu{} scaling relations because they are more fundamental.
The scaling relation for $\nu_{\rm max}$ is based on the assumption that $\nu_{\rm max}$ in solar-like oscillators is a fixed fraction of $\nu_{ac}$\citep{1991ApJ...368..599B,1995A&A...293...87K}. The acoustic cutoff frequency ($\nu_{ac}$) is the frequency above which acoustic waves are not reflected at the surface and it depends on the effective temperature and the surface gravity as $\nu_{ac} \propto g T_{\rm eff}^{-0.5}$. 
%The physical meaning 
Regarding the $\Delta \nu$ scaling relation, the large frequency separation $\Delta \nu$ in the asymptotic approximation equals the inverse of the sound travel time through the star, which results in $\Delta \nu \propto \sqrt{M/R^3}$ \citep{1986ApJ...306L..37U}. The $\Delta \nu$ value is hence proportional to the square root of the mean density ($\overline{\rho}$) of the star.

\subsection{Correcting the $\nu_{\rm max}$ Scaling Relation}

We calculated the scaling $\nu_{\rm max}$ using model $T_{\rm eff}$ and $\log g$, and used the same fitting method to constrain it. We refer to this estimated scaling $\nu_{\rm max}$ as $\nu_{\rm max, SR}$. Note that we did not use the measured $\nu_{\rm max}$ as an input constraint in the fitting, and hence the determination of $\nu_{\rm max, SR}$ is independent of the observed value.   

\begin{figure}
\centering
\includegraphics[scale=0.45]{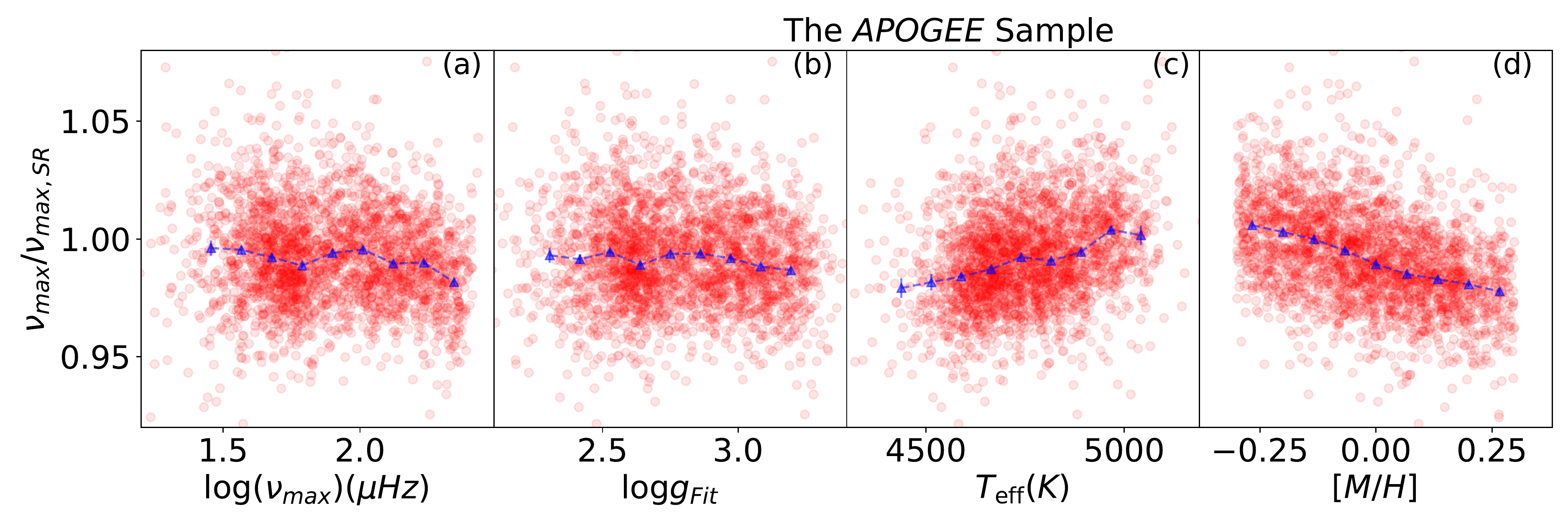}
\includegraphics[scale=0.45]{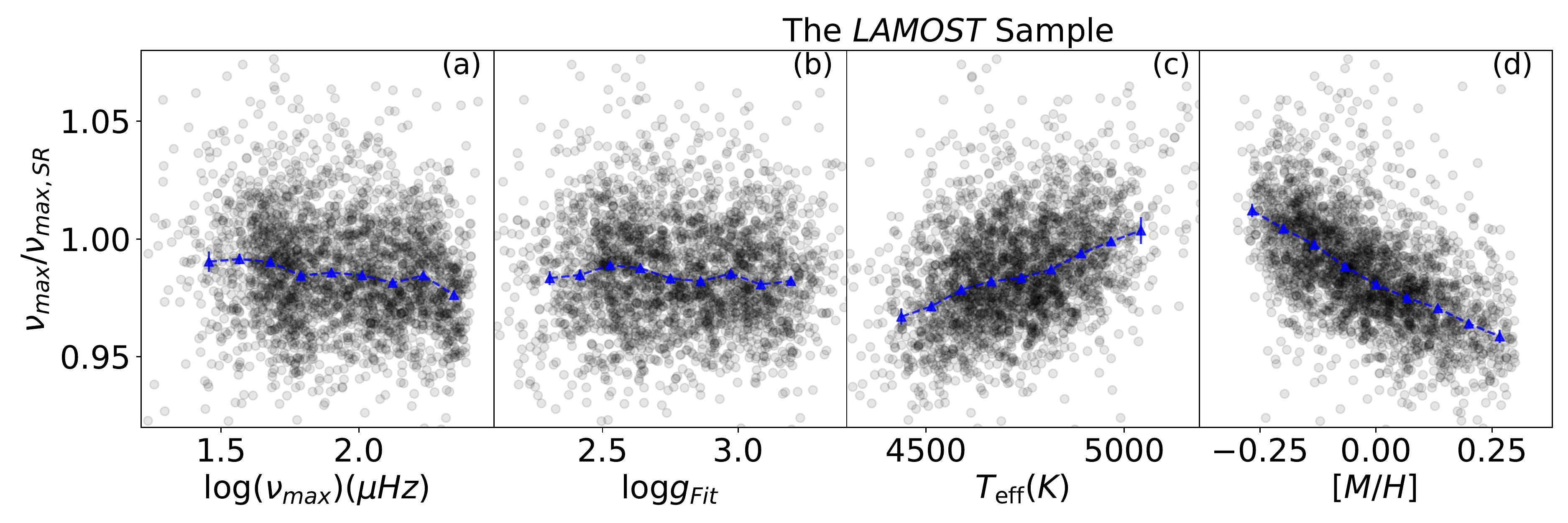}
\caption{Dependencies of systematic offsets in $\nu_{\rm max}$ ($\nu_{\rm max}$/$\nu_{\rm max,SR}$) on $\nu_{\rm max}$, $T_{\rm eff}$, $\log g$,and [Fe/H]. The \apogee{} sample is presented on the top and the \lamost{} sample is shown at the bottom. Filled triangle is the median of binned data and the error bar indicates the the standard error of the median, which is calculated as the standard deviation over the number of data point in a bin ($\sigma /n$). 
\label{fig:offsetnumax}}
\end{figure}

From the comparison between observed $\nu_{\rm max}$ and $\nu_{\rm max,SR}$, as demonstrated in Figure \ref{fig:offsetnumax}a, we find no significant systematic offset. However, the scatter (the standard deviation is $\pm$2.5\%) is much larger the typical observed uncertainty, indicating potential dependencies on other stellar parameters or that \numax{} uncertainties are substantially underestimated. 
We therefore investigated the correlations with $\log g$, $T_{\rm eff}$, and {\it [M/H]} in Figures \ref{fig:offsetnumax}b, \ref{fig:offsetnumax}c, and \ref{fig:offsetnumax}d. The reason to consider the metallicity is that it could impact near-surface properties of solar-like oscillators, as suggested by hydrodynamic simulations \citep[e.g.][]{2014MNRAS.445.4366T}. 
%Previous studies \citep[e.g.][]{2019MNRAS.486.4612B} on seismic dwarfs also suggested to include a metallicity term in the scaling relation formulae.
%
We see that the $\nu_{\rm max}$ offset does not obviously depend on $\log g$, but clearly correlates with $T_{\rm eff}$ and [M/H]. The same trend is seen in both the \apogee{} and \lamost{} samples. 
From the perspective of physics, the correlations with the effective temperature may be real because the thermal structures at the near-surface region are very different between RGB and main-sequence stars, although there is no theory that quantifies the difference in \teff{}-\numax{} relations between the two types of stars. However, we suspect that this correlation is caused by the systematic offsets in the spectroscopic measurements because the dependency on \teff{} is slightly stronger for the \lamost{} data than for the \apogee{} data. 
The same applies for the metallicity, in that the correlation is likely a combination of the physical and the instrumental effects and we cannot separate them because of the lack of theoretical predictions. We will leave this as an open question for future studies.

To correct for the systematics discussed above, we added a metallicity term to the $\nu_{\rm max}$ scaling relation. We calibrated the revised \numax{} scaling relation with model-based $\log g$, observed \numax{}, \teff{}, and [M/H]. We used the \textsc{Scipy curve\_fit} module to do the fits and obtained corrected $\nu_{\rm max}$ scaling relations as follows:  
\begin{equation}\label{eq:new_numax1}
  \frac{\nu_{\rm max}}{\nu_{\rm max,\odot}} = \frac{g_{\rm Fit}}{g_{\odot}} \left ( \frac{T_{\rm eff}}{T_{\rm eff, \odot}} \right )^{-0.459} \left (10^{\rm [M/H]_{}}  \right)^{-0.022} (\apogee)\\\\
\end{equation}  
  and
\begin{equation}\label{eq:new_numax2}  
  \frac{\nu_{\rm max}}{\nu_{\rm max,\odot}} =  \frac{g_{\rm Fit}}{g_{\odot}} \left ( \frac{T_{\rm eff}}{T_{\rm eff, \odot}} \right )^{-0.421} \left (10^{\rm [M/H]_{}}  \right)^{-0.039} (\lamost).\\
\end{equation}
We obtained the same power law for the $g$ term as the original relation, but the exponent of $T_{\rm eff}$  is slightly different from the original ($-0.5$).
As discussed above, the difference does not necessarily mean that the RGB stars have a different \teff{}--\numax{} relation from the dwarfs. For instance, when we considered a global temperature shift of $-70K$ in the \apogee{} sample, as found by \citet{2019MNRAS.486.3569H}, the fit gave a very similar power law to the original scaling relation.    
{ The additional term of [M/H] is a secondary term in the formulae, which causes a $\sim$4\% change to the \numax{} value when [M/H] varies from -0.3 to 0.3 dex. 
Note that the \lamost{} [M/H] values are based on the [Fe/H] and [$\alpha$/Fe] measurements. The choice of formula may bring some systematic uncertainties and we hence tested the fits with another widely-used formula, namely [M/H] = [Fe/H] + [$\alpha$/Fe]. The [M/H] values from this formula are higher than those calculated with Eq.~\ref{eq:mh} by 0.02 dex, on average, which is significantly smaller than the typical observed uncertainty. Fitting the scaling relation with these [M/H] values returns very similar results to Eq.~\ref{eq:new_numax2}. The exponent is $-0.427$ for the \teff{} term and 0.040 for the [M/H] term. The similarity indicates the choice of [M/H] formula does not significantly affect the new scaling relation. }

% The systematic offset in $\nu_{\rm max}$ ranges in $\pm \sim$4\% obviously correlated to $T_{\rm eff}$, and [Fe/H]. We obtained a corrected $\nu_{\rm max}$ scaling relation as with a random uncertainty of 2.5\%. 
% \begin{equation}\label{eq:sc_numax}
%   \frac{g_{\rm Fit}}{g_{\odot}} = \frac{\nu_{\rm max}}{\nu_{\rm max,\odot}} \left ( \frac{T_{\rm eff}}{T_{\rm eff, \odot}} \right )^{0.534} \left (10^{\rm [Fe/H]}  \right)^{0.025} (\apogee)\\\\
% \end{equation}  
%   and
% \begin{equation}\label{eq:sc_numax}  
%   \frac{g_{\rm Fit}}{g_{\odot}} = \frac{\nu_{\rm max}}{\nu_{\rm max,\odot}} \left ( \frac{T_{\rm eff}}{T_{\rm eff, \odot}} \right )^{0.481} \left (10^{\rm [Fe/H]}  \right)^{0.038} (\lamost)\\
% \end{equation}

\subsection{The $\Delta \nu$ Scaling Relation}

\begin{figure}
\centering
\includegraphics[scale=0.45]{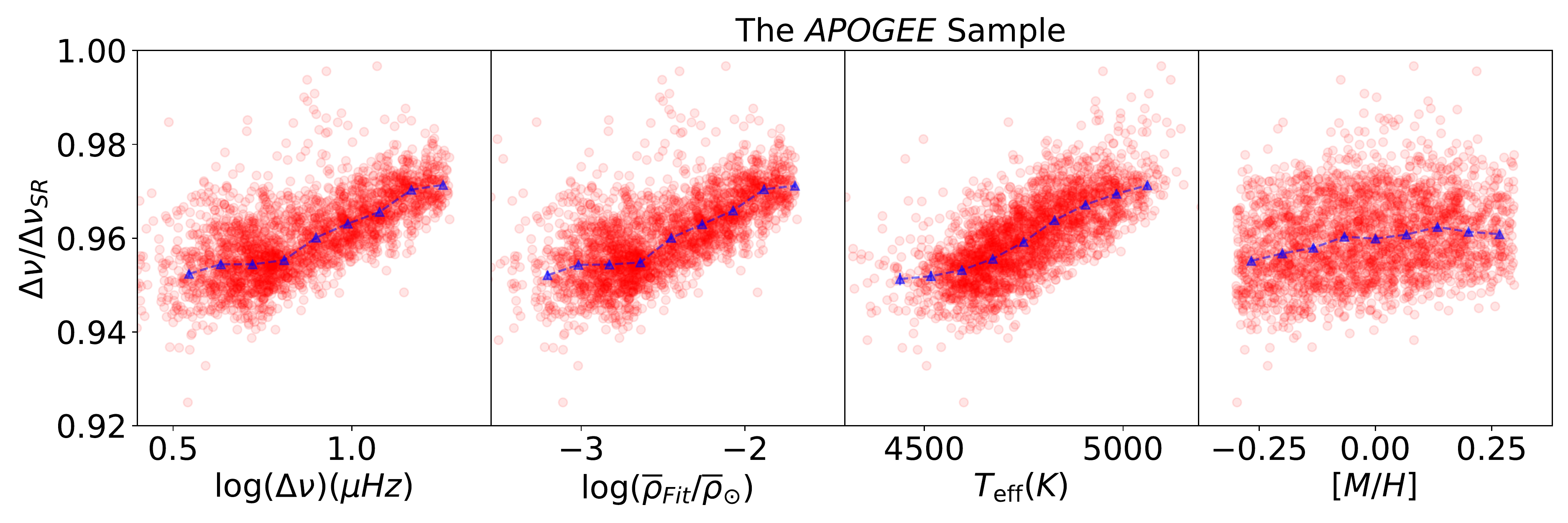}
\includegraphics[scale=0.45]{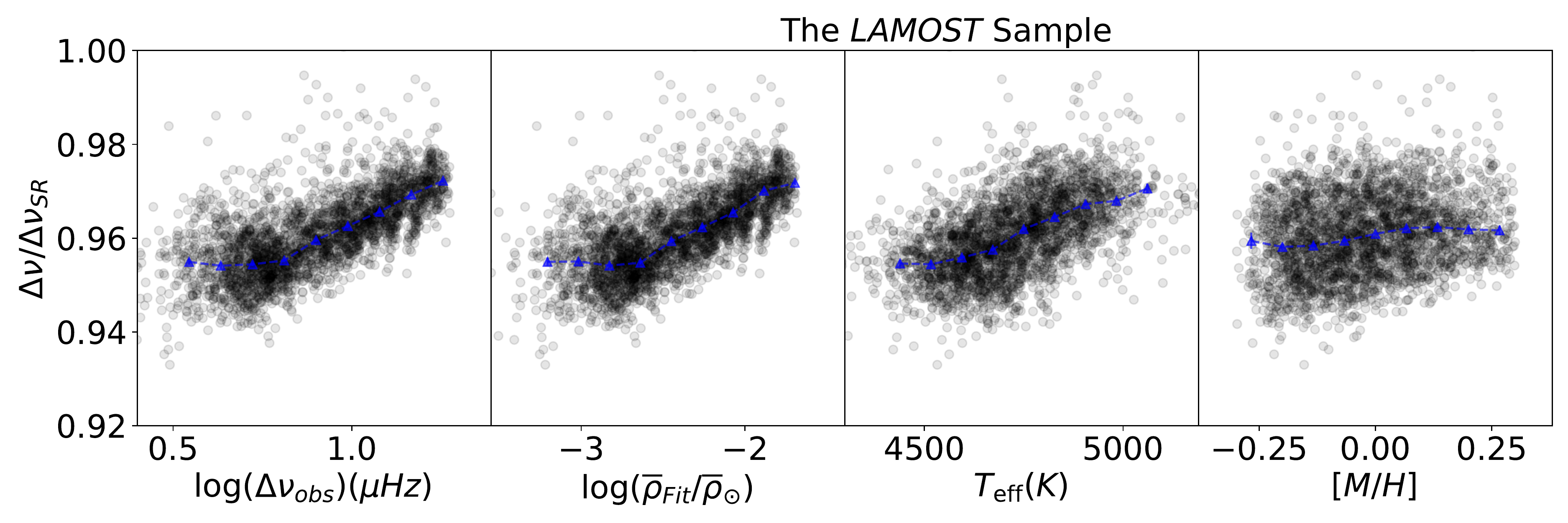}
\caption{Same as Figure \ref{fig:offsetnumax} but for \Dnu{}.    
\label{fig:offsetdnu}}
\end{figure}

We studied the $\Delta \nu$ scaling relation following the same strategy. 
We used the mean density of models to calculate the large frequency separation from the scaling relation (Eq.~\ref{eq:sc-Dnu}) and constrained it based on model fitting. We refer to this as $\Delta \nu_{\rm SR}$.
We first examined the systematic difference between observed $\Delta \nu$ and $\Delta \nu _{\rm SR}$. Note that the observed \Dnu{} is from \citet{yuj++2018-16000-rg} rather than from individual radial modes.
As shown in Figure \ref{fig:offsetdnu}a, there is a $\sim$4\% deviation on average, and the offset varies from $-6$\% to $-2$\%. We then inspected the dependencies of $\Delta \nu$/$\Delta \nu_{\rm SR}$ on the mean density, effective temperature, and metallicity, as demonstrated in Figures~\ref{fig:offsetdnu}b, \ref{fig:offsetdnu}c, and \ref{fig:offsetdnu}d. 

The ratio clearly correlates with mean density and also with the effective temperature, and slightly depends on the metallicity. 
It is not clear which parameter has the true effect because these three parameters are correlated on RGB. 
As for its physical significance, we note that $\Delta \nu$ is determined by the mean density rather than by any surface properties. As pointed out by \citet{2012MNRAS.419.2077M}, the differences in internal temperature (hence sound speed) distributions could cause a noticeable difference in $\Delta \nu$ between different type of stars, despite them having the same mean density. Their comparison between hydrogen-shell-burning and helium-core-burning stars with similar mean densities in $\kepler$ open clusters presented a $\sim$3.3\% difference in average $\Delta \nu$.
As a consequence, the difference in $\Delta \nu$--$\rho$ scaling relation between RGB and main-sequence stars is expected because of their different internal structures. 
The correlations with effective temperate and metallicity arise because these two surface properties relate to the mean density: 
for stars on RGB, the effective temperature strongly correlates with mass, and hence with mean density, at given radius. Moreover, high-resolution spectroscopic surveys have shown a clear \teff{}--[M/H] relation on the RGB \citep[e.g. Fig.2][]{2017ApJ...840...17T}. For this reason, we retained the functional form of \Dnu{} scaling relation and did not include the effective temperature or the metallicity in the new scaling relation.    

We calibrated the $\Delta \nu$--$\rho$ scaling relation using observed $\Delta \nu$ and modeling-inferred mean density ($\rho_{\rm Fit}$), obtaining the following results:
\begin{equation}\label{eq:new_dnu}
  \frac{\Delta\nu}{\Delta\nu_{\odot}} = \left( \frac{\overline{\rho}_{\rm Fit}}{\overline{\rho}_{\odot}}  \right) ^{0.507}.\\
\end{equation} 
Note that this function is for both the \apogee{} and \lamost{} samples because there is no systematic differences in inferred mean densities between the two at the sample level. Compared with the scaling for \numax, the scaling for \Dnu{} has a significantly larger systematic offset and the corrections are especially important for RGB stars. The correction to the \Dnu{} scaling relation also matches the evolution-dependent systematic offsets found in previous studies \citep[e.g.][]{2016ApJ...822...15S}. We see that the power law exponent relating the large separation and the mean density for RGB stars is larger than that for dwarfs. This causes \Dnu{} offsets (in fractional terms) to grow exponentially with decreasing \Dnu{}, and hence leads to larger systematic offsets for more evolved stars.
We also note that a departure from the 0.5 exponent in the \Dnu{} scaling relation has also been seen in $\delta$~Scuti stars \citep[e.g.,][]{Garcia-Hernandez++2015,Bedding++2020}.

\begin{figure}
\centering
\includegraphics[scale=0.4]{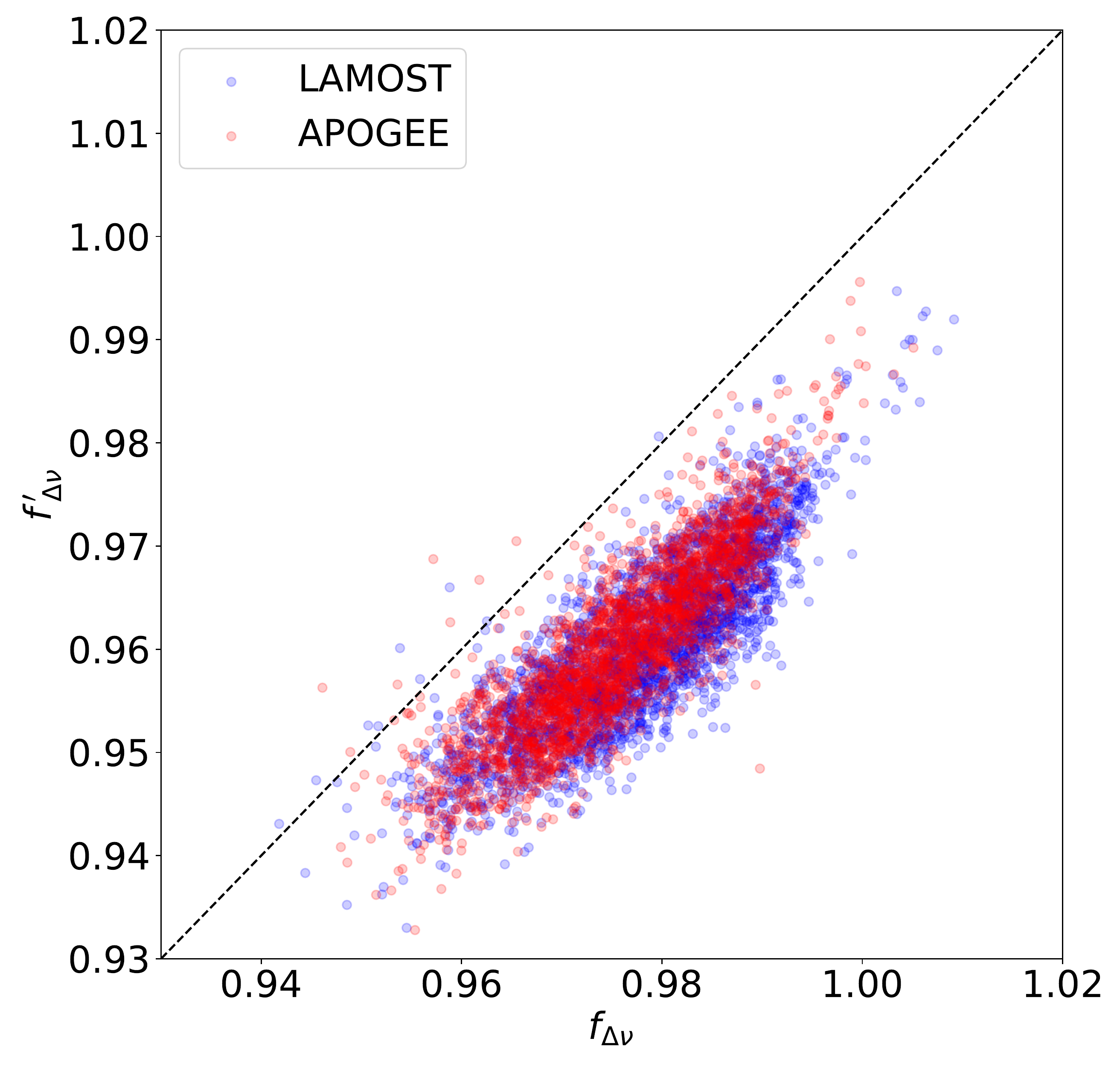}
\caption{The comparison between the two correction factors to the \Dnu{} scaling relation ($f_{\Delta\nu}$ and $f\,'_{\Delta\nu}$) for the \apogee{} sample. Here, $f\,'_{\Delta\nu}$ can be taken as the an improved version of $f_{\Delta\nu}$ due to an additional correction for the surface term.
\label{fig:compare_fdnu}}
\end{figure}

Our correction to the \Dnu{} scaling relation is similar to previous studies by \citet{2011ApJ...743..161W} and \citet{2016ApJ...822...15S}, who checked whether the \Dnu{} from fitting the model frequencies matched the value from the scaling the model density. This work is an improvement on that earlier work in that our corrections include the surface correction to model frequencies. 
It would be useful to investigate how much the surface term affect the corrections \Dnu{} scaling relation.
To do this, we followed the method described by \citet{2011ApJ...743..161W} and \citet{2016ApJ...822...15S} to calculate the correction factor, $f_{\Delta\nu}$. In line with their notation, we defined our \Dnu{} corrections as $f\,'_{\Delta\nu}$ = \Dnu/\Dnu$_{\rm SR}$.
With this notation, we can re-write Eq. \ref{eq:new_dnu} as
\begin{equation}
      \frac{\Delta\nu}{\Delta\nu_{\odot}} = f\,'_{\Delta\nu} \left( \frac{\overline{\rho}_{\rm Fit}}{\overline{\rho}_{\odot}}  \right )^{0.5}.\\
\end{equation}
Here, $f\,'_{\Delta\nu}$ can be taken as an improved version of $f_{\Delta\nu}$, with an additional correction for the surface term.
Figure \ref{fig:compare_fdnu} shows the comparison between $f_{\Delta\nu}$ and $f\,'_{\Delta\nu}$ of the best-fitting models for all stars. The differences are roughly uniform and the average differences are 1.6\% and 1.4\% for the \apogee{} and \lamost{} samples, respectively. 
If we adopt a 1.5\% difference in the \Dnu{} correction factor and transfer this difference to the mass and radius determinations, the surface term will cause a systematic offset of $\sim$6\% for the mass and $\sim$3\% for the radius, which are highly significant. We hence conclude that the surface term is worthy of attention in the corrections to the \Dnu{} scaling relation.

\begin{figure}
\centering
\includegraphics[scale=0.45]{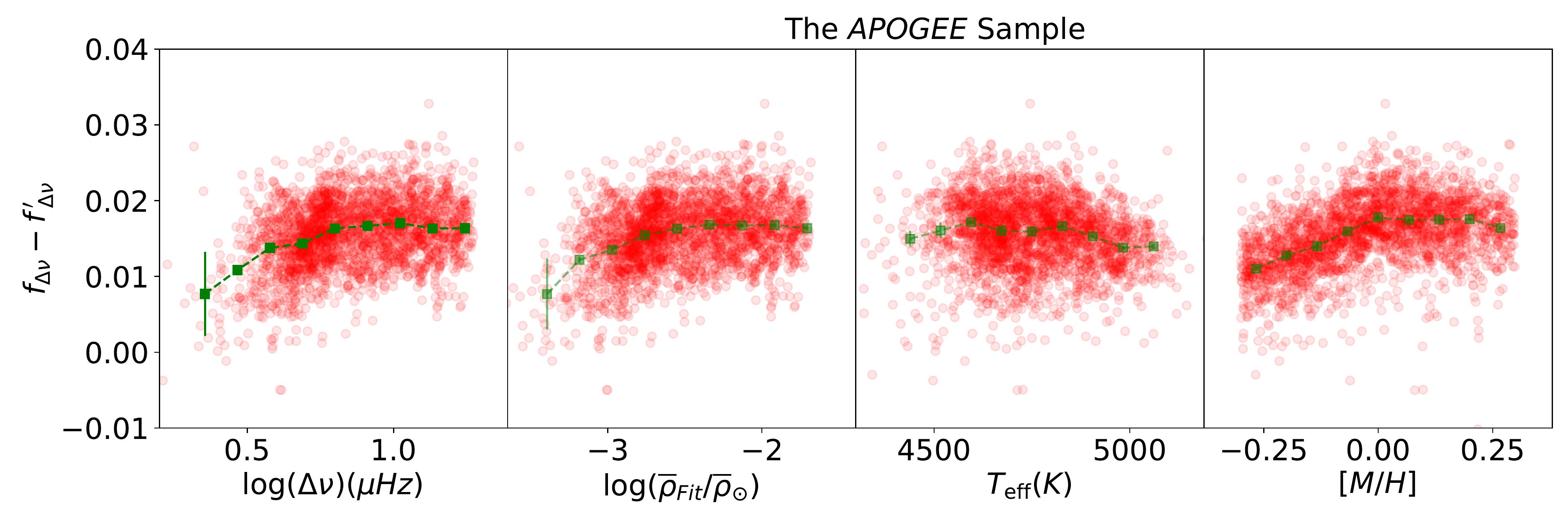}
\caption{The difference in the two correction factors to the \Dnu{} scaling relation ($f\,'_{\Delta\nu}$ - $f_{\Delta\nu}$) as a function of four stellar parameters for the \apogee{} sample. Green squares indicates the rolling median of the data. Results for the \lamost{} sample (not shown)  are similar.
\label{fig:fdnu_corre}}
\end{figure}

The surface term depends on properties of near-surface layers and we hence inspected its correlations with the mean density, effective temperature, and metallicity. As shown in Figure~\ref{fig:fdnu_corre}, the difference between the two factors decreases with \Dnu{}, implying that the surface term is smaller for more evolved stars. This trend is consistent with 3D convection simulations of the surface effect on the RGB \citep[e.g.][]{2017MNRAS.466L..43T}. The effective temperature also has an effect on the surface term according to those simulation results, but it is not obvious in our sample because the effective temperature range on RGB is fairly narrow. We also notice a rough correlation with the metallicity. This is expected because the metallicity can also change the convection properties at near surface \citet[e.g.][]{2015A&A...573A..89M}.

%
%In case there are potential dependency the effective temperature or the metallicity, we examine the residuals of the fit and find no obvious correlation to any of the two parameters.  

% : a $\sim$3.3\% difference in mean $\Delta \nu$ between stars on the RGB and in the core helium burning phase.

% The accuracy of $\Delta \nu$ scaling relation has been studied and corrected with several different approaches. Using $\Delta \nu$ from a linear fit to theoretical frequencies of radial modes, \citet{2011ApJ...743..161W} and \citet{2016ApJ...822...15S} found deviations from the scaling relation in models and they computed correction factor to $\Delta \nu$ ($f_{\Delta\nu}$). The values of $f_{\Delta\nu}$ correlate to the effective temperature and slightly depends on the model input metallicity for RGB stars. 
% However, these correlations could be partly because of the surface term, which also relates to effective temperature and metallicity but is not corrected in both studies. 
%
%

\subsection{Corrected Scaling Relations for Red Giants}
With corrected \numax{} and \Dnu{} scaling relations (Eq. \ref{eq:new_numax1}, \ref{eq:new_numax2} and \ref{eq:new_dnu}), we derived new scaling relations for determining stellar masses and radii. For the $\apogee$ sample they are: 
\begin{equation}\label{eq:nsc1a}
   \frac{M_{\rm Fit}}{M_{\odot}} = \left ( \frac{\nu_{\rm max}}{\nu_{\rm max,\odot}} \right )^{3.0}\left ( \frac{\Delta\nu}{\Delta\nu_{\odot}} \right )^{-3.944}\left ( \frac{T_{\rm eff}}{T_{\rm eff, \odot}} \right )^{1.377}\left (10^{\rm [M/H]} \right)^{0.066}, 
\end{equation}
and
\begin{equation}\label{eq:nsc2a}
   \frac{R_{\rm Fit}}{R_{\odot}} = \left ( \frac{\nu_{\rm max}}{\nu_{\rm max,\odot}} \right )\left ( \frac{\Delta\nu}{\Delta\nu_{\odot}} \right )^{-1.972}\left ( \frac{T_{\rm eff}}{T_{\rm eff, \odot}} \right )^{0.459}\left (10^{\rm [M/H]} \right)^{0.022}.
\end{equation}
For the $\lamost$ sample they are: 
\begin{equation}\label{eq:nsc1b}
   \frac{M_{\rm Fit}}{M_{\odot}} = \left ( \frac{\nu_{\rm max}}{\nu_{\rm max,\odot}} \right )^{3.0}\left ( \frac{\Delta\nu}{\Delta\nu_{\odot}} \right )^{-3.944}\left ( \frac{T_{\rm eff}}{T_{\rm eff, \odot}} \right )^{1.263}\left (10^{\rm [M/H]} \right)^{0.117}, 
\end{equation}
and
\begin{equation}\label{eq:nsc2b}
   \frac{R_{\rm Fit}}{R_{\odot}} = \left ( \frac{\nu_{\rm max}}{\nu_{\rm max,\odot}} \right )\left ( \frac{\Delta\nu}{\Delta\nu_{\odot}} \right )^{-1.972}\left ( \frac{T_{\rm eff}}{T_{\rm eff, \odot}} \right )^{0.421}\left (10^{\rm [M/H]} \right)^{0.039}.
\end{equation}
It worth noting that the above relations are only valid for RGB stars within a parameter range of $T_{\rm eff} = 4400$ to 5200 K,  $\nu_{\rm max} = 10$ to 270 $\mu$Hz, and ${\rm [M/H]} = -0.3$ to 0.3 dex. We suggest caution when applying them outside these parameter ranges. 

\begin{figure}
\includegraphics[scale=0.32]{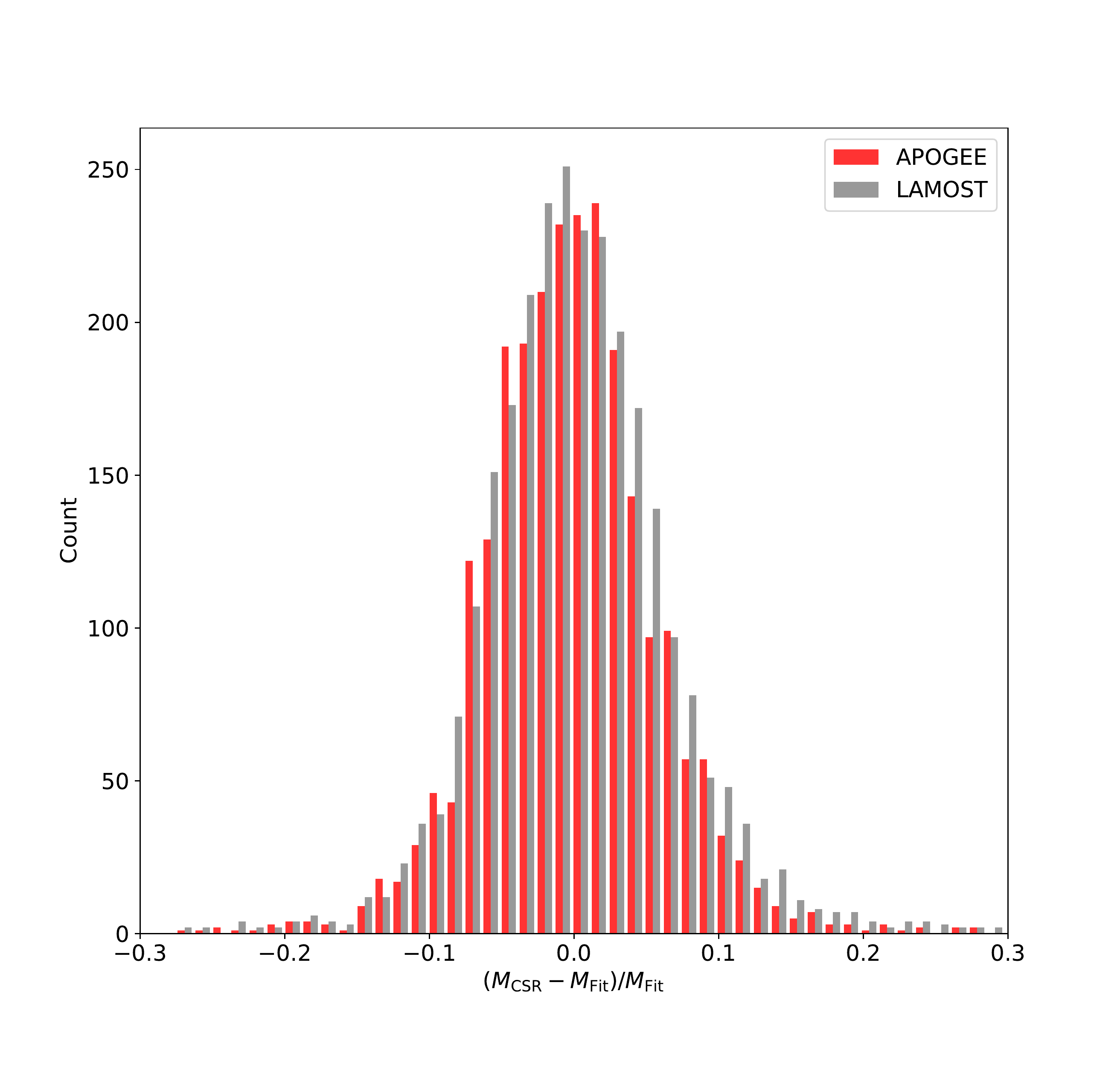}
\includegraphics[scale=0.32]{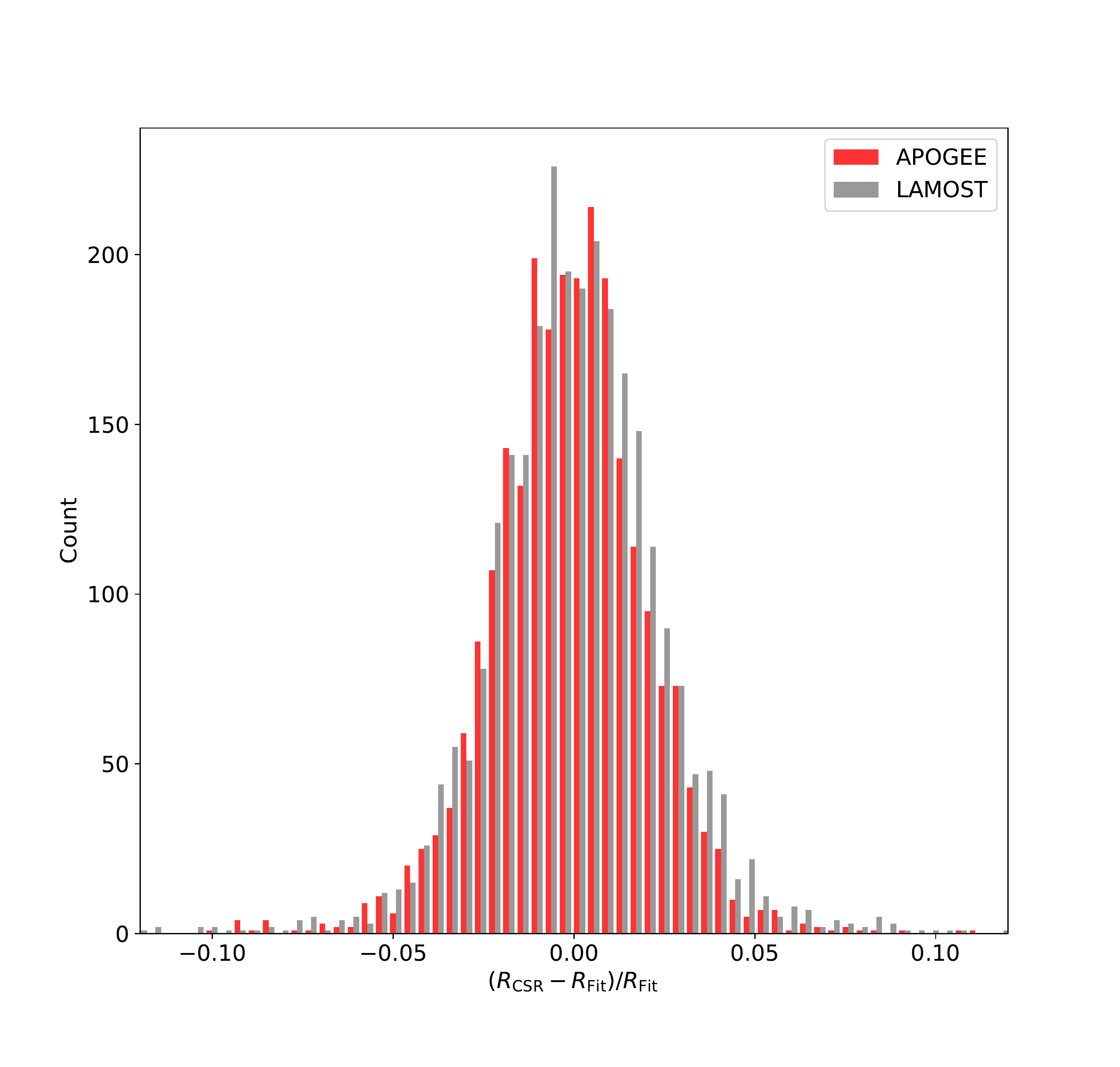}
\caption{Differences between modeling-inferred results and corrected scaling relations (represented by the subscript `CSR'). Both density distributions center around zero with a standard deviation of $\sim5\%$ for mass and $\sim2\%$ for radius.
\label{fig:corr_results}}
\end{figure}

As a test, we computed masses and radii with the new scaling relations and compared them with model-determined values. We illustrate the distributions of the residuals in Figure \ref{fig:corr_results}. The distributions center at zero with a Half-Width-at-Half-Maximum of $\sim$5\% for the mass and of $\sim$2\% for the radius. These are consistent with the typical estimated uncertainties, which are 4.5\% and 1.7\% for mass and radius.

Because the new scaling relations are based on modeling solutions, some systematic uncertainties are expected. We considered reasonable ranges for the helium fraction and the mixing-length parameter, so the systematic effects from these two inputs are covered in the estimated uncertainty. However, the choices of other input physics could bias the results to some extent. These include the solar composition, the opacity, the atmosphere boundary condition, mass-loss rate, and atomic diffusion. The systematic offset, according to comparisons between seven different pipelines for \kepler{} main-sequence stars \citep{2017ApJ...835..173S}, is about $\sim$2\% for the mass and $\sim$1\% for the radius on average. Systematic offsets related to input physics in our work could be different because internal structures of RGB stars are very different from those of dwarfs. Future studies are required to properly estimate the model systematic uncertainties.

We calibrated scaling relations for the $\apogee$ and $\lamost$ samples separately because of the systematic differences in their measurements. Given the large number of $\apogee$ and $\lamost$ spectra, the new scaling relations should be useful for future studies. 
It is also worth considering whether to apply these new scaling relations to stars observed by other spectroscopic surveys. 
To test this, we applied the \apogee{} formulae to the \lamost{} data.
As determined from Figure \ref{fig:hrd}, we uniformly increased the \lamost{} \teff{} by 44K, and calibrated the \lamost{} [M/H] using the linear function as 0.20[M/H] + 0.023. We used calibrated \lamost{} \teff{} and [M/H] in Eq.~\ref{eq:nsc1a} and \ref{eq:nsc2a} to calculate masses and radii. Comparing with model-based determinations, we found good consistency: the average difference was zero with a spread of $\pm$1.3\% for the mass and $\pm$0.4\% for the radius. This result indicates that the new scaling relations can be applied with spectroscopic data from other sources, provided proper calibrations to the effective temperature and the metallicity are implemented.

% It could be noted that the metallicity term in Eq. \ref{eq:sc13} or \ref{sc:23} is relatively secondary term. Given the metallicity range of our sample ranges from -0.3 to 0.3, the term raises a 3\% variation for mass and a 0.7\% variation for radius. The values are much smaller than what is illustrated in Figure \ref{fig:scaling_correlation}. This is because the matallicity strongly correlates to the effective temperature with given $\nu_{\rm max}$ for the evolutionary stages of our sample. An example can be found in XXX. This is to say, the metallicity effect is mostly included in the temperature effect in the fitting. Thus, we could simplify the above four-terms version by removing the matellicity term. This gives us three-terms formulae which are very useful for stars without metallicity measurements. We fit the three-terms formulae to our modeling results and obtain

\section{Conclusions}\label{sec:conclusion}

We studied the solar-like oscillations of 3,642 \kepler{} RGB stars. We measured radial mode frequencies and used them to characterise stars with detailed modeling, including correction for the surface effect. 
We provide model-based estimates for mass, age, radius, and surface gravity for the full sample.
For five red giants in eclipsing binaries, we found good consistency with masses and radii determined from dynamical modeling. We also showed that systematic differences between the \apogee{} and the \lamost{} measurements of effective temperature and metallicity do not significantly affect the modeling solutions at the sample level because the fits are dominated by seismic mode frequencies.

We used our results to study the systematic offsets in the widely-used \numax{} and \Dnu{} scaling relations. 
We did not find a significant systematic offset in the \numax{} scaling relation. However,  
a clear dependency on observed metallicity was seen, and we hence added a metallicity term to the scaling relation. We used the observed \numax{}, \teff{}, [M/H], and model-inferred $\log g$ to calibrate our revised \numax{} scaling relation for RGB stars. We noted that the correction to the \numax{} scaling relation depends on the selection of spectroscopic measurements.
For the \Dnu--density relation, our results showed that the scaling relation overestimates the \Dnu{} value by $\sim$4\% on average. We calibrated a new \Dnu{}--$\rho$ relation for RGB stars based model-inferred $\overline{\rho}$ and observed \Dnu, and showed that the surface correction plays a very important part in this calibration.

Combining our results, we derived revised scaling relations for determining masses and radii for RGB stars. The new relations can characterise stars with precision of $\sim$5\% in mass and $\sim$2\% in radius. We suggest that the new scaling relations can be applied to stars observed by other spectroscopy surveys, provided any differences in the effective temperature and metallicity scales are properly calibrated.

\acknowledgments
We gratefully acknowledge the {\em Kepler} team, whose efforts made these results possible.
This work has received funding from the European Research Council (ERC) under the European Union’s Horizon 2020 research and innovation programme (CartographY GA. 804752). 
This work is supported by the Joint Research Fund in Astronomy (U2031203) under cooperative agreement between the National Natural Science Foundation of China (NSFC) and Chinese Academy of Sciences (CAS) and also by NSFC grants 12090040 and 12090042.
We also gratefully acknowledge support from the Australian Research Council (grant DP210103119), and from the Danish National Research Foundation (Grant DNRF106) through its funding for the Stellar Astrophysics Center (SAC).

\bibliography{aalib,astrobib_yaguang}{}
\bibliographystyle{aasjournal}

\end{document}